\documentclass[twocolumn]{aastex63}
\usepackage{soul}
\usepackage{hyperref}

\newcommand{\fbedo}{FB80\_D1}
\newcommand{\fbedt}{FB80\_D2}
\newcommand{\fbtdt}{FB20\_D2}
\newcommand{\fbtdo}{FB20\_D1}

\newcommand{\dthr}{D3}
\newcommand{\dtwo}{D2}
\newcommand{\done}{D1}

\newcommand{\lzuni}{kpc km/s}
\newcommand{\lz}{$L_z$}
\newcommand{\vrotsig}{$v_{rot}/\sigma^*$}
\newcommand{\euni}{$\rm{km}^2/\rm{s}^2$}

\newcommand\gastro{{\tt GASTRO}}
\newcommand\gasoline{{\tt GASOLINE}}
\newcommand{\msun}{\mbox{$\,\mathrm{M_{\odot}}$}}

\newcommand{\kms}{\mbox{$\>{\rm km\, s^{-1}}$}}
\submitjournal{ApJ}

\shorttitle{{\tt GASTRO} library I: accreted stars}
\shortauthors{Amarante et al.}
\graphicspath{{./}{figures/}}

\begin{document}

\title{\gastro\, library I: the simulated chemodynamical properties of several GSE-like stellar halos}

\correspondingauthor{Jo\~ao A. S. Amarante}
\email{joaoant@gmail.com}

\author[0000-0002-7662-5475]{Jo\~ao A. S. Amarante}\altaffiliation{Visiting Fellow at UCLan}
\affiliation{Institut de Ciències del Cosmos (ICCUB), Universitat de Barcelona (IEEC-UB), Martí i Franquès 1, E08028 Barcelona, Spain}
\affil{Jeremiah Horrocks Institute, University of Central Lancashire, Preston, PR1 2HE, UK}


\author[0000-0001-7902-0116]{Victor P. Debattista}
\affil{Jeremiah Horrocks Institute, University of Central Lancashire, Preston, PR1 2HE, UK}
\author[0000-0002-0740-1507]{Leandro {Beraldo e Silva}}
\affil{Jeremiah Horrocks Institute, University of Central Lancashire, Preston, PR1 2HE, UK}
\affil{Department of Astronomy, University of Michigan, 1085 S. University Ave., Ann Arbor, MI 48109, USA}
\author[0000-0003-3922-7336]{Chervin F. P. Laporte}
\affiliation{Institut de Ciencies del Cosmos (ICCUB), Universitat de Barcelona (IEEC-UB), Martí i Franquès 1, E08028 Barcelona, Spain}
\author[0000-0003-3523-7633]{Nathan Deg}
\affil{Department of Physics, Engineering Physics, and Astronomy, Queen's University, Kingston ON K7L~3N6, Canada}
%



\begin{abstract}
The Milky Way stellar halo contains relics of ancient mergers that tell the story of our Galaxy's formation. Some of them are identified due to their similarity in energy, actions, and chemistry, referred to as the ``chemodynamical space", and are often attributed to distinct merger events. It is also known that our Galaxy went through a significant merger event that shaped the local stellar halo during its first billion years. Previous studies using $N$-body only and cosmological hydrodynamical simulations have shown that such single massive merger can produce several ``signatures" in the chemodynamical space, which can potentially be misinterpreted as distinct merger events. Motivated by these, in this work we use a subset of the \gastro\, library which consists of several SPH+$N$-body models of a single accretion event in a Milky Way-like galaxy. Here, we study models with orbital properties similar to the main merger event of our Galaxy and explore the implications to known stellar halo substructures. We find that: \textit{i.} supernova feedback efficiency influences the satellite's structure and orbital evolution, resulting in distinct chemodynamical features for models with the same initial conditions, \textit{ii.} very retrograde high energy stars are the most metal-poor of the accreted dwarf galaxy and could be misinterpreted as a distinct merger, \textit{iii.}  the most bound stars are more metal-rich in our models, the opposite of what is observed in the Milky Way, suggesting a secondary massive merger, and finally \textit{iv.} our models can reconcile other known apparently distinct substructures to a unique progenitor. 
\end{abstract}




\section{Introduction}\label{sec:intro} 
In the $\Lambda$CDM paradigm, the build up process of the stellar halo of a Milky Way-mass galaxy includes the accretion of hundreds of luminous satellite galaxies into the main progenitor \citep[e.g.][]{searle,bullockjohnston2005, rey-starkenburg2022}. This view is supported by observations of high redshift galaxy mergers \citep[e.g.][]{patton+2002,lin+2004} and also by numerical simulations of galaxy formation \citep[e.g.][]{moore+1999,benson+2000,Newton_2018}. The accreted stars will then build the host galaxy's stellar halo, and signatures in the chemodynamical space of these stars help trace back the accretion history of now disrupted satellite galaxies. \par
Beyond  30 kpc in Galactocentric radius, the Milky Way's (MW) outer stellar halo exhibits a large amount of past accretion events in the form of stellar streams \citep[e.g.][]{belokurov+2006, shipp+2018, ibata+2021} as well as a number of ongoing accretion events such as the Sagittarius dwarf \citep[e.g.][]{ibata+1994,johnston1995, majewski+2003, laporte+2019} and the Large and Small Magellanic Clouds \citep[e.g.][]{kallivayalil+2006,besla+2007}. While most of the stellar streams are detected as spatial overdensities in photometric surveys with relatively little spectroscopic follow-up, the inner stellar halo is currently being studied in greater detail thanks to the advent of large spectroscopic surveys and the \textit{Gaia} mission. \par   
With the increasing number of stars with full 6D phase-space information and measured chemical abundances, there are currently dozens of unambiguously\footnote{Some substructures are re-identified and given different names.} identified substructures in the MW's inner stellar halo \citep[for an overview, see e.g.][]{yuan+2020,  helmi2020, naidu+2020, limberg+2021, ruiz-lara+2022, malhan+2022}. In principle, different mergers should be distinguishable in the dynamical space, e.g. energy, angular momentum, and actions, in a slowly evolving potential \citep[as shown in e.g.][]{johnston+1996,helmi1999}. This has been explored extensively in order to find signatures of merger event(s) with pure $N$-body \citep[e.g.][]{jean-baptiste+2017,koppelman+2020,naidu+2021_nbody} and cosmological hydrodynamic simulations \citep[e.g.][]{wu+2022}. These have shown that ({\it i.}) different merger events can overlap in dynamical space even if they started with different orbit configuration and ({\it ii.}) single merger events can leave their imprints on a large range of energy and actions. These results lead to the question whether all substructures have a separate origin or whether several are the relics of the same merger event. \par
Since the early 2000's there has been growing evidence of a significant merger event that shaped the MW's inner stellar halo ($r_{gc}\lesssim 30$ kpc during the Galaxy's first Gyrs, \citealt{chiba&beers2000,gilmore2002, meza2005, nissen-schuster2010, deason+2013}). This event, recently dubbed the ``\textit{Gaia}-Sausage-Enceladus" (GSE), has been confirmed \citep{belo2018, helmi2018} thanks to the exquisite data from \textit{Gaia} satellite \citep{gaia1,gaiahr}. The GSE has a net rotational velocity close to zero and a large radial velocity dispersion, in contrast to the more isotropic velocity distribution of the metal-poor, [Fe/H]$<-2$, stellar halo \citep{belo2018}. The merger is estimated to have happened during the first three Gyr of the MW's formation (e.g. \citealt{gallart+2019}, but see \citealt{donlon+2020} for an alternative time of the merger). Its median [Fe/H] ranges from $\sim-1.45$ to $-1.17$ depending on the spectroscopic survey and sample selection \citep[e.g.][]{fernandez-alvar+2017,amarantea,feuillet+2020, das+2020, naidu+2020, bonifacio+2021, buder+2022}.  Its stellar mass is estimated to be in the range $10^{8.5}$-$10^{9.7}$ \msun (e.g. \citealt{vincenzo+2019, feuillet+2020, naidu+2020, hasselquist+2021, limberg+2022}) which would correspond to a dark matter halo mass of $\sim 2-7\times 10^{11}$\msun\, \citep[e.g.][]{moster+2013}.\par
Besides explaining the observational features, simulations can also guide us to understand better the GSE merger. For instance, pure $N$-body models \citep[][]{koppelman+2020,naidu+2021_nbody} constrained the GSE dwarf to be on a retrograde orbit (with respect to the MW's disk rotation) with low to intermediate circularity, and with a low inclination angle between its orbital plane and the MW's disk plane. GSE-like mergers have also been identified/studied in cosmological simulations of MW-like galaxies \citep[e.g.][]{fattahi+2019,mackereth+2019,bignone+2019,grand+2020, elias+2020, dillamore+2022}. For example, \citet{fattahi+2019} selected MW-like galaxies in Auriga \citep{grand+2017} and found that about 1/3 of them have a highly anisotropic, metal-rich, [Fe/H] $\approx -1$, stellar halo, which becomes isotropic at lower metallicities, i.e., just as is observed in the MW. In the simulations, these stellar halos are mostly assembled during the first Gyrs of the galaxy evolution. Moreover, the typical stellar mass of such mergers is of the order of $10^9-10^{10}$\msun, similar to what is estimated in the EAGLE simulations, $10^{8.5}-10^{9}$\msun\citep[][]{mackereth+2019}\par
Given its relatively massive and extended nature, how is the GSE's debris dispersed in the MW? Could the GSE merger event result in more than one of the substructures observed in the chemodynamical space of the local stellar halo? The former question has been explored with pure $N$-body models by \citet{koppelman+2020} and \citet{naidu+2021_nbody} who showed that such a merger extends over a large range in the energy-angular momentum space. In this work, we will tackle the second question exploring a suite of $N$-body+smoothed particle hydrodynamics (SPH) models of a GSE-like dwarf merging into a MW-like galaxy. Due to the galaxy formation model that self-consistently follows the growth of the host galaxy and its merging satellite -- which cannot be done with a pure $N$-body model -- we are able to self-consistently study the chemical evolution of the merging dwarf and its chemodynamical properties at the end of the simulation, i.e. when the accreted stars are dynamically relaxed. In these simulations, a MW-like galaxy grows from the cooling of a hot gas halo and experiences a merger with a dwarf galaxy in a fine-tuned orbit. Thus, these simulations are intended to reproduce the MW in a tailored manner. 

In this paper, we will show the structural evolution of different satellites during their merger process into the host galaxy and how this depends on the strength of the supernova feedback of the models. Also, due to the discy nature of dwarfs, we will demonstrate that a GSE-like merger event produces a rich behavior in the accreted stars' chemodynamical space. This has important consequences for the known MW substructures as we are able to link some of them dynamically and chemically to the GSE merger. \par
The paper is organized as follows: Section~\ref{sec:gastro} introduces the \gastro\, library of simulations and the subset of the models analyzed in this work. In Section~\ref{sec:evolution} we describe the evolution of the dwarfs during the merger process. We then explore the chemodynamical signatures of the GSE-like merger event in Section~\ref{sec:results}. We discuss the observational implications of our models in section Section~\ref{sec:implication} and list our conclusions and final remarks in Section~\ref{sec:conc}. 
\section{GASTRO library}\label{sec:gastro}
The \textit{Gaia-EncelAdus-Sausage Timing, chemistRy and Orbit} (\gastro)\footnote{\url{amarante.netlify.app/GASTRO}} library is a suite of $N$-body + SPH simulations exploring the parameter space for GSE-like accretion events into the MW and its impact on the formation history of the MW. In this work, we use a subset of the \gastro\, library to explore the chemodynamical trends of the accreted stars. A complete description of the \gastro\, library will be presented in a future paper. Throughout this paper we analyze 4 models (and 4 extra ones in the appendix) where a GSE-like galaxy merges  with a MW-like galaxy. In Section \ref{sec:gasoline} we describe the input physics of the code used to run the models. In Section \ref{sec:hostIC} and \ref{sec:dwarfIC}, we describe the initial conditions of the host galaxy and the GSE-like dwarf galaxy models, respectively. Finally, in Section \ref{sec:mergerIP} we describe the set up of the dwarf's initial orbital parameters.
\subsection{GASOLINE input physics}\label{sec:gasoline} 
All the models are evolved with \gasoline\, \citep{wadsley} which implements SPH into the $N$-body tree code {\tt PKDGRAV} \citep{pkdgrav}. The gas phase uses the metallicity dependent cooling method of \citet{shen2010}. We set a pressure floor on gas particles $p_{floor}=3G\epsilon^2\rho^2$, where $G$ is the universal gravitational constant, $\epsilon$ is the softening length and $\rho$ is the gas particle's density \citep{agertz+2009}. Star formation begins wherever the temperature drops below 15,000 K and the density exceeds 1 $cm^{-3}$ in a convergent flow \citep{stinson}. The feedback by supernova explosions follows the blastwave implementation of \citet{stinson}. In this work, we present models which couple either 20\% or 80\% of the $10^{51}$ erg per supernova as thermal energy injected into the interstellar medium. In isolated galaxy simulations, low feedback models have been shown to produce geometrical and chemical thin and thick disks \citep{clarke+2019, beraldo+2020}, as well as a kinematically hot thick disk population \citep{amaranteb}, very similar to those observed in the MW. Turbulent diffusion of the gas \citep{shen2010} is also included allowing the gas to mix, which reduces the scatter in the age-metallicity relation \citep{pilkington+2012}.
The integration base time step is set to $\Delta t = 10$ Myr and is refined such that $\delta t = \Delta t/2^2 < \nu \sqrt{\epsilon/a_g}$, where $a_g$ is the acceleration at a particle's position and the refinement parameter $\nu = 0.175$. The opening angle, which determines when to open a branch in the $N$-body tree code, is set to $\theta=0.7$. The time step of gas particles satisfies the further condition $\delta t = \nu_{courant}h/[(1+\alpha)c + \beta \mu_{max}]$, where the Courant number for SPH criteria $\eta_{courant} = 0.4$, $h$ is the SPH smoothing length set over the nearest 32 particles, $\alpha$ and $\beta$ are the linear and quadratic viscosity coefficients and $\mu_{max}$ is the maximum viscous force measured between the gas particles \citep{wadsley, springel2010}. \par
Each star particle represents a single stellar population following a Miller-Scalo initial mass function \citep{miller-scalo1979} with a initial mass 1/3 of the initial parent gas particle. Once a gas particle loses 80\% of its mass, its mass is distributed amongst the nearest neighboring gas particles, which avoids having an inert population of low-mass gas particles. Supernova type II and Ia yields of Oxygen and Iron are taken from \citet{raiteri+1996}. Supernova type II and type Ia rates are determined using Padova stellar lifetimes \citep[as in][]{raiteri+1996} with the latter taking into account a binary evolution model. 

\subsection{Milky Way-like galaxy set up}\label{sec:hostIC} 
The initial conditions set for the host galaxy are the same as in the isolated model used by our group which describes several chemo-kinematic properties of the Milky Way: the geometrical and chemical thin/thick disks \citep{clarke+2019, beraldo+2020}, the ``Splash" population \citep{amaranteb}, the existence of a significant old thin disk population in the Solar neighborhood \citep[][with a higher mass resolution of the same model]{beraldo+2021} and the bulge chemistry (Debattista et al. in prep.). In the following we summarize the initial conditions of the host galaxy. \par
The initial conditions are characterized by a spherical hot gas corona embedded in a dark matter halo with a Navarro-Frenk-White \citep[NFW]{NFW} density profile with virial radius $r_{200}\approx 200$ kpc, mass of $10^{12}\text{M}_\odot$, central density $2.4\times10^6\text{M}_\odot{\rm kpc}^{-3}$ and scale radius of $30$ kpc. Dark matter (DM) particles come in two mass flavors of $10^6$\msun\, and $3.5\times10^6$\msun\, inside and outside 200 kpc, respectively -- in total the model has $10^6$ DM particles. The gas corona follows the same radial density profile but constitutes only 10\% of the total mass of the system. There are $10^6$ gas particles each with mass of $1.4\times10^5$ \msun. The gas has an initial net rotation, with spin parameter $\lambda=0.065$ \citep{bullock+2001}, cools via metal-line cooling (see Section \ref{sec:gasoline}) and settles into a disk. The softening length used for DM and gas is 100 pc and 50 pc, respectively. The $N$-body+SPH nature of the models allows star particles to form self-consistently from the cold gas, inheriting the softening length and chemistry of the parent gas particle.  \par
\subsection{Dwarf galaxy set up}\label{sec:dwarfIC}
\begin{table*}
 
 \begin{center}
 \begin{tabular}{ccccccc}
 \hline
 Model & Mass [\msun] & $\langle {\rm [Fe/H]} \rangle$ & $\Delta {\rm [Fe/H]}/\Delta r$ [dex/kpc]& $\langle {\rm [O/Fe]} \rangle$ & $\Delta {\rm [O/Fe]}/\Delta r$ [dex/kpc] & VRHE fraction (\%)\\
  \hline
  \fbedo & $3.15\times10^8$ & $-1.14$  &  $-0.17$ & 0.2 & $0.039$ & $3.72$\\
  \fbtdo & $8.97\times10^8$ &  $-0.65$ &  $-0.29$ & 0.31 & $0.034$ & $1.5$ \\
  \fbedt & $5.53\times10^8$ & -1.08 &  $-0.18$ & 0.23 & $-0.015$ & $2.06$\\
  \fbtdt & $1.18\times10^9$ & -0.77 &  $-0.25$ & 0.28 & $0.016$ & 1.33\\
  
 \hline
 \end{tabular}
  \caption{Properties of the dwarf in each model. The mass, $\langle {\rm [Fe/H]}\rangle$, and $\langle {\rm[O/Fe]}\rangle$ are measured at the epoch of the merger. The [Fe/H] and [O/Fe] gradients are measured before the dwarfs reach their pericenter, i.e. when they are still rotationally supported. Finally, the very retrograde high energy (VRHE) stellar mass fraction of the satellite is measured at $t=10$ Gyr. }
  \label{tab:models}
  \end{center}
\end{table*}

Currently the \gastro\, library explores seven different dwarf galaxy models with varying initial gas and dark matter distributions. In this paper, we show the results for two dwarf galaxies (and a third one in Appendix B) and Table \ref{tab:models} summarizes some of their properties. The variations in their initial conditions have important consequences for the star formation, and thus to the chemical evolution, of each dwarf. Throughout the paper, we will use the words ``dwarf" and ``satellite" interchangeably, always referring to the less massive galaxy. 
\par
The dwarfs \done, \dtwo\, and \dthr\, (shown only in the appendix) are generated using the \textsc{GalactICS} code \citep{kuijken-dubinski1995, widrow-dubinsky2005, widrow+2008}. This particular iteration of the code allows for the generation of equilibrium, exponential gas disks. For a full description of the new features we refer the reader to \citet{deg+2019}. In our runs, all dwarfs are initialized with a DM and gaseous disk, with $10^5$ and $2\times10^4$ particles respectively. These are parameterized by NFW halos with a scale velocity of $200 \kms$ and are truncated at $50$ kpc. The gas disks of the three dwarfs are exponentials built with an initial kinematic temperature of $1000\,\textrm{K}$ and have a scale radius of $4$ kpc. \done\, and \dthr\, DM halos have gas disc exponential scale radii of $1$ kpc, while \dtwo\, has a scale radius of $5$ kpc.
The initial gas mass for each dwarf is $1.4\times10^{9}\msun$, $2.75\times10^{9}\msun$ and $5.5\times10^{8}\msun$ for \done, \dtwo\, and \dthr\, respectively. 

\subsection{Merger impact parameters}\label{sec:mergerIP}
In this paper, we focus on two orbital scenarios involving two merger timescales: one at $\sim 3$ Gyr and another at $\sim 2$ Gyr. Throughout the text we will refer to this timescale as the ``merging time", which is when the dwarf is completely disrupted. We set the dwarf at a distance from the center of the host galaxy of 200 kpc and 150 kpc, for the former and the latter. In both scenarios the initial circularity of the orbit, $\eta\equiv L_z/L_c(E)$, is set to 0.3, where $L_z$ is the z-component of the angular momentum and $L_c(E)$ is the angular momentum of circular motion for an orbit of energy $E$ in the Galactic plane. Finally, all dwarf-models have their orbit inclination relative to the host galaxy's mid-plane set to $15^{\circ}$. \par
We note that \citet{naidu+2021_nbody}, using pure $N$-body models, found that the optimal initial orbital conditions were: $r_{gc} = 150$ kpc, orbit circularity 0.5 and orbit inclination $15^{\circ}$. While we show here scenarios with the same initial $r_{gc}$ and orbit inclination, our models differ in orbit circularity. This particular choice is based on preliminary analysis of the effects on the host galaxy disk depending on the initial orbit circularity of the satellite. Nonetheless, in  Figure \ref{fig:appen:enlz} we show models in which satellites \dtwo\, and \dthr\, are placed on orbits with $\eta =0.5$ and it is clear that they also have qualitatively similar trends in the Energy-$L_z$ space as the models discussed in the main text.\par
In the next section, we study the evolution of the dwarf properties in each model. Given our present focus on the properties of the stellar halo, the study of the impact of the GSE-like mergers on the evolution of the Galactic disk is left to a separate paper. 
\section{Evolution of the models}\label{sec:evolution}
In this paper, we study four models, where each is set up with the initial conditions of the host galaxy + a dwarf model. We use dwarfs \done\, and \dtwo\, while also varying the feedback strength. The low (high) feedback models are \fbtdo\, and \fbtdt\, (\fbedo\, and \fbedt)  in which 20\% (80\%) of the supernova energy is injected back to the interestellar medium (see Section \ref{sec:gasoline}).
The models are evolved for 10 Gyrs and we save snapshots every 0.1 Gyr. All our models' host galaxy, regardless of the dwarf merger or feedback prescription used, end up roughly with the same final total stellar mass of $\approx 6\times10^{10} \msun$ ($\approx 2\times10^6$ star-particles). The total rotation curve of the models is in good agreement with the MW's. For instance, its total circular velocity at the Solar radius is $v_{c\odot}\approx 260$ km/s and has an approximately flat-profile toward the outer disk. The evolution of the dwarf's structure prior and during the merger are more pertinent to the present work so we describe them in this section.\par
\subsection{Dwarf galaxy orbital decay}
\begin{figure*}
    
    \includegraphics[width=\textwidth]{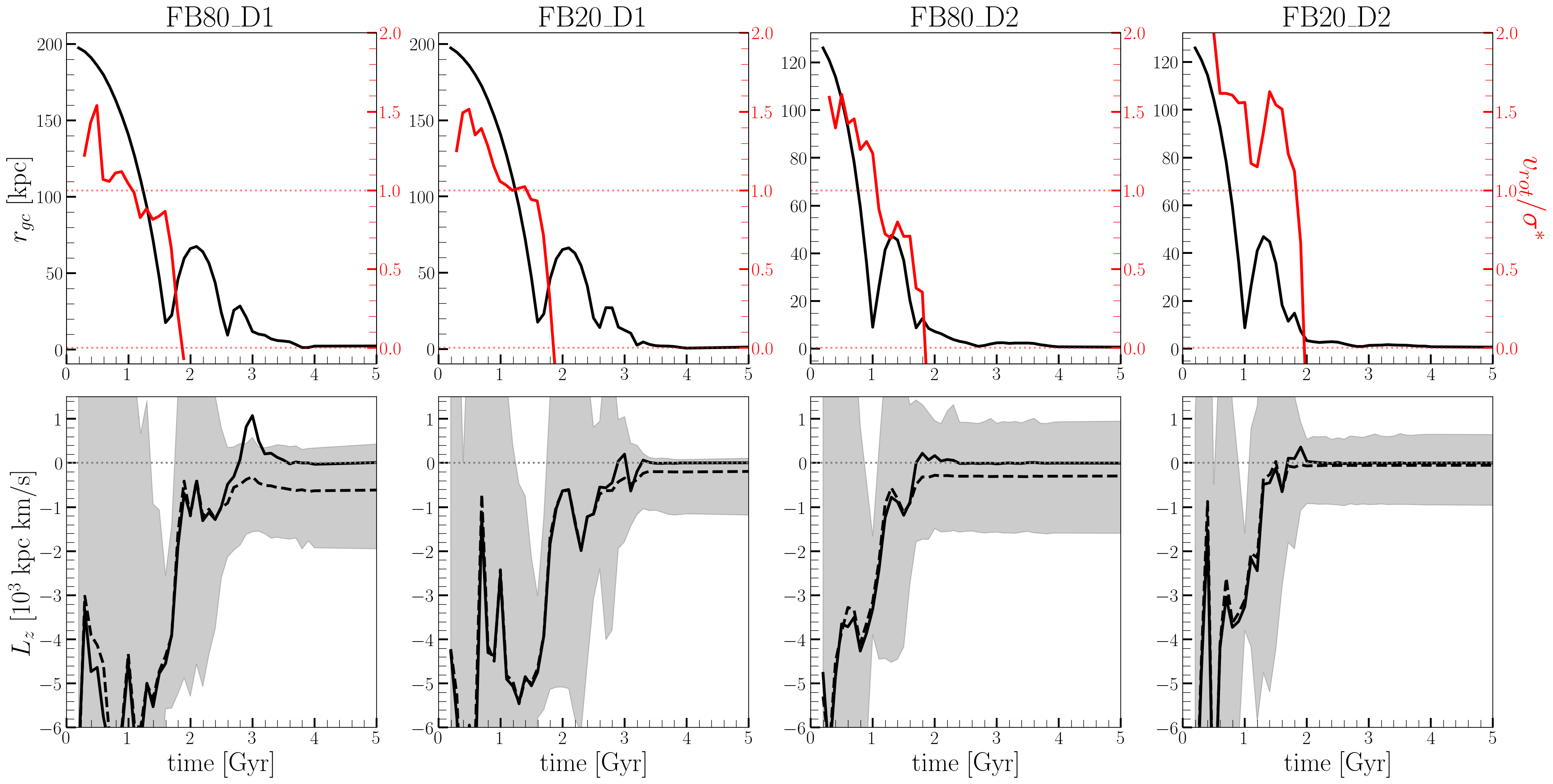}
    \caption{\textbf{Top row:} The evolution of the distance between the dwarf's and the host galaxy centers represented by the black line with the scale shown on the left-axis. We also show the dwarf's $v_{rot}/{\sigma^*}$ evolution (red line and scale shown on the right axis). See text for details. \textbf{Bottom row:} The median $L_z$ evolution of the dwarf's center of mass and all accreted stars shown as solid and dashed lines respectively. While the center of mass ends up with  $L_{z,orb}  = 0$ \lzuni, the overall distribution show signs of bulk retrograde motion, except in model \fbtdt, shown in the last column. The shaded area corresponds to the 5\% and 95\% percentiles of the $L_z$ distribution. }
    \label{fig:lzevol}
\end{figure*}

The dwarfs \done\, and \dtwo\, start at a distance of $r_{gc}=200$ kpc and $r_{gc}=150$ kpc, respectively, from the center of the host galaxy. In all cases, the satellite's velocity is such that its initial orbit circularity is $\eta=0.3$. As each model evolves and form stars, we track the orbital evolution of the dwarf with respect to the host by following its center of mass, COM, distance, $r_{com}$, and the vertical component of its orbit angular momentum, $L_{z,orb}$, measured with respect to the host galaxy's plane. The top row of Figure \ref{fig:lzevol} shows the evolution of $r_{com}$ for all the models as the solid black line. The dwarfs in models \fbtdo\, and \fbedo\, have merging timescales of $t\approx3$ Gyr, and have two apocenter passages. In contrast, models \fbtdt\, and \fbedt\, have comparatively shorter merging timescales of $t\approx2$ Gyr after only one apocenter. \par
The satellite's orbital decay can also be studied through the evolution of $L_{z,orb}$ as shown by the solid line in the bottom row of Figure \ref{fig:lzevol}. The dwarfs \done\, and \dtwo\, start with $|L_{z,orb}| \approx 5.8$ and $4.5\times 10^3$ \lzuni, respectively. Their COM loses absolute angular momentum during the orbital decay until the orbit is radialized at $L_{z,orb}\approx 0$ \lzuni. The evolution of the median $L_z$ of all the dwarf's stars (measured with respect to the host galaxy) is represented by the dashed black line. The shaded area corresponds to the interpercentile range 5-95\%. The accreted stars in the high feedback models, \fbedo\, and \fbedt, end up with a lower median $L_z$ compared to their low feedback counterparts. Moreover, the final $L_z$ distribution is broader in the higher feedback models. This is caused by feedback directly impacting the evolution of the internal structure of the dwarf galaxies, which in turn affect the final angular momentum distribution. Finally, the median $L_z$ follows the $L_{z,orb}$ until the last pericenter at which point the dwarf is completely disrupted.
We will explore how the feedback impacts the internal structure of the dwarf galaxies in the following section.
\subsection{Dwarf internal structural evolution}\label{sec:dwprofile}
In the previous section, we showed that the strength of the feedback does not affect the dwarfs' COM orbital evolution, but influences the final $L_z$ distribution of the accreted stars. In order to understand this result, we now proceed to explore the effect of feedback on the internal structure of the satellites. We start by checking the evolution of \vrotsig, where $v_{rot}$ and $\sigma^*$ are the dwarf's mean rotational velocity and the total velocity dispersion, respectively, measured in the dwarf's frame. If \vrotsig$>1$ the object is rotationally supported. When a satellite is orbiting a massive host galaxy, such as the ones explored here, \citet{kazantzidis+2011} demonstrated that rotationally supported dwarfs can be tidally stirred, by losing their rotational angular momentum, during gravitational interactions with the host galaxy. \par
We study the evolution of \vrotsig\, of the dwarfs for the stars within $1$ kpc of its center -- the trend also holds at different radii. This is shown in the top row of Figure \ref{fig:lzevol} as the solid red line. Models using \done, which is the less massive and radially extended model, show similar evolution of \vrotsig. Before the first pericenter, the center of the dwarfs are mildly rotationally supported, becoming pressure supported after the first pericenter passage. \par
Models accreting dwarf \dtwo\, show a different \vrotsig\, evolution depending on the feedback strength. Before the first pericenter, both models have similar \vrotsig$\approx1.5$. However, soon after, the high feedback model, \fbedt, loses all of its rotational support due to the interaction with the host galaxy, already reaching \vrotsig$<1$. On the other hand, the low feedback model, \fbtdt, recovers part of its internal rotational support due to a stronger starburst in its central parts. As the dwarf approaches the final pericenter, it becomes pressure supported, i.e. \vrotsig$\approx0$.\par    %
\begin{figure*}
    \includegraphics[width=\textwidth]{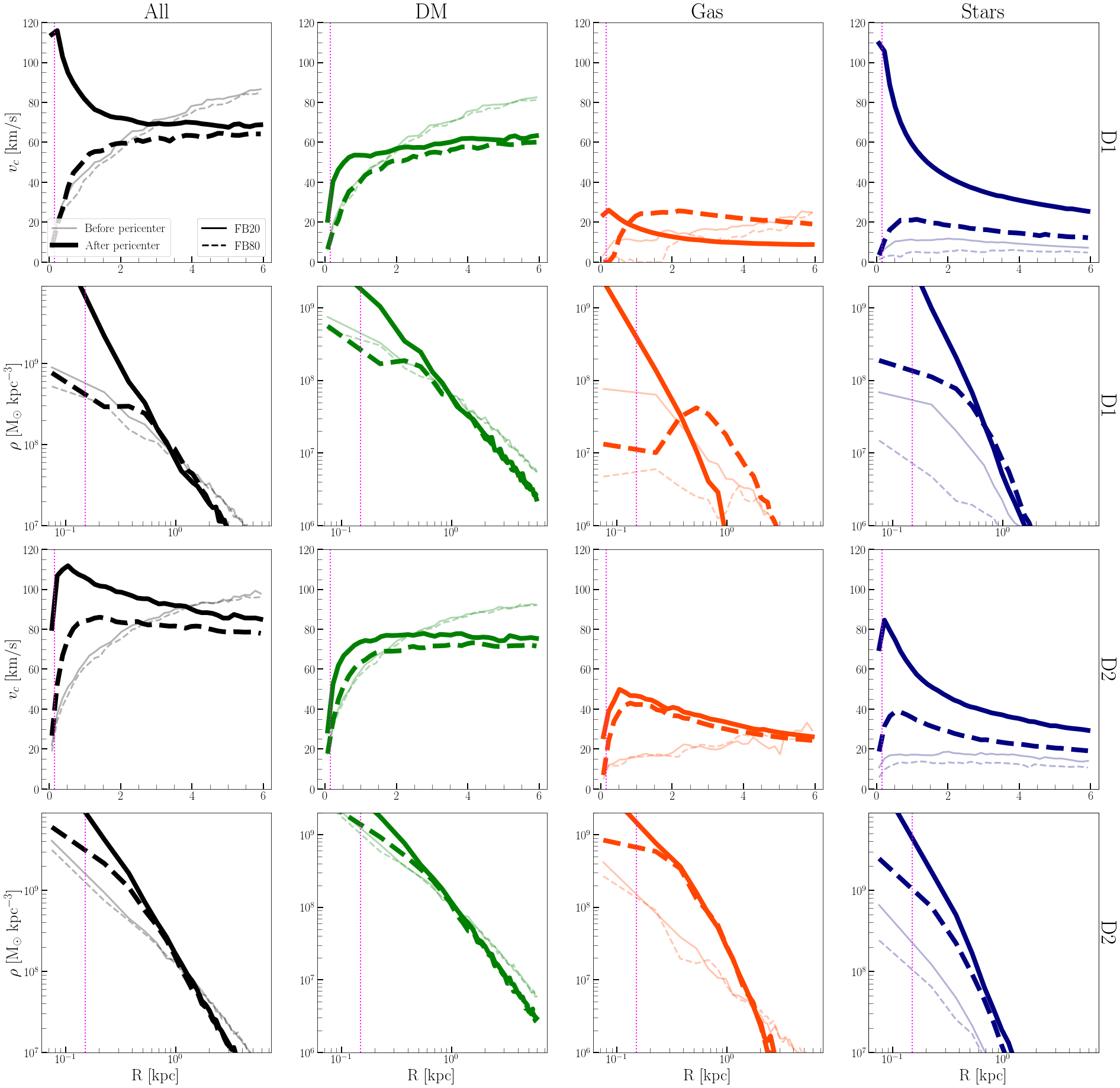}
    \caption{The circular velocity and mass density radial profiles for \done\, (\dtwo) are shown in the first (last) two rows. From left to right the total, dark matter, gas and stellar mass profiles for the low and high feedback models are shown as solid and dashed lines, respectively. Before the first pericenter passage, the profiles for different feedback models, shown as the lighter colors, do not show significant difference. After the first pericenter and before the dwarf is completely disrupted, the low feedback models have a more centrally concentrated mass profile, whereas the high feedback models maintain a shallower distribution. The magenta dashed line delimits $r =0.3$ kpc, which is three times the star-particle softening length.}
    \label{fig:dwarfprofile}
\end{figure*}
The effect of different feedback models can also be seen in the dwarf's rotation curve and density profile evolution. The first and second rows of Figure \ref{fig:dwarfprofile} show the evolution of satellite \done's\, circular rotation curve, $v_c$, which is calculated based on the total mass within its midplane, and spherical mass density profile, $\rho$, respectively. The dark matter, gas and stellar mass content are shown from left to right columns. Before the first pericenter passage, at $t=1.4$ Gyr, both feedback models have similar behavior, as shown by the lighter colors lines. The dark lines show the profiles before total disruption at $t=2.4$ Gyr. At this point, the mass profiles are drastically different depending on the feedback strength. The DM, gas, and stellar mass distributions in the low feedback model are more centrally concentrated. This effect is enhanced by the tidal interaction of the dwarf with its host: the dwarf center is compressed and, in the lower feedback model, together with its centrally concentrated gas density, allows a higher star formation rate \citep[since high feedback suppresses star formation, e.g.][]{hopkins+2012, faucher+2013} in the central parts, leading to a more centrally concentrated stellar mass distribution compared to the high feedback model. Finally, the higher stellar mass content leads to adiabatic contraction of the dark matter to the center \citep{blumenthal+1986, gnedin+2004}, making its distribution more centrally concentrated in the low feedback regime.   \par
The last two rows show the profile for dwarf \dtwo, at $t=0.9$ Gyr and $t=1.4$ Gyr, i.e. before and after the first pericenter passage, with light and dark colors respectively. This dwarf shows the same trends with feedback as discussed above, with the difference that its stellar content is more radially extended in comparison with dwarf D1. The fact that low feedback models have a centrally concentrated mass distribution allows their stellar content to lose more $L_z$, becoming more radialized compared to the higher feedback counterpart (as shown in Figure \ref{fig:lzevol}, see also a detailed discussion on the radialization of mergers in \citealt{vasiliev+2022}).\par
As previously mentioned, models with single merger events have already been extensively explored with $N$-body only simulations. Now as we add more complexity to these models, by adding gas physics and star formation, we start to see extra dependence on the adopted input physics. For instance, so far we have shown that the feedback strength influences the satellite's internal structure and, consequently, affects the satellite stars' final $L_z$ distribution. In the following section we will also discuss the effect of feedback on the chemical evolution of the dwarfs.

\subsection{The dwarf galaxy chemistry}\label{sec:dwchem}
As the models evolve and form stars self-consistently, there is a natural chemical evolution of the host and satellite galaxies. In particular for the dwarf, which is the main focus in the present paper, this will reflect on the final chemodynamical configuration of the accreted stars (see Section \ref{sec:results}) and impact the interpretation of the known substructures in the MW (see Section \ref{sec:implication}). Therefore, we now explore the chemical evolution of the dwarfs. \par
We start by measuring the [Fe/H] and [O/Fe] radial abundance profiles for each model. While the [Fe/H] evolution is robust for different stellar yields models, the $\alpha$-abundance, here tracked by [O/Fe], varies strongly with stellar yields and is in general used in a more qualitative\footnote{For instance, it is still possible to distinguish $\alpha$-poor and $\alpha$-rich stellar population.} sense \citep[as shown in][]{buck+2021}. We measure the gradients by fitting a linear function to the radial abundance profile before the dwarf's first pericenter, i.e. at $t=1.4$ Gyr and $t=0.9$ Gyr for dwarfs \done\, and \dtwo\, respectively. At this time, the dwarf is still rotationally supported and has not been stirred by the tidal interactions (see Section \ref{sec:dwprofile}). The measured gradients are presented in Table \ref{tab:models}. \par
\begin{figure*}
    
    \includegraphics[height=\dimexpr
  \textheight-3\baselineskip-\parskip-.2em-
  \abovecaptionskip-\belowcaptionskip\relax]{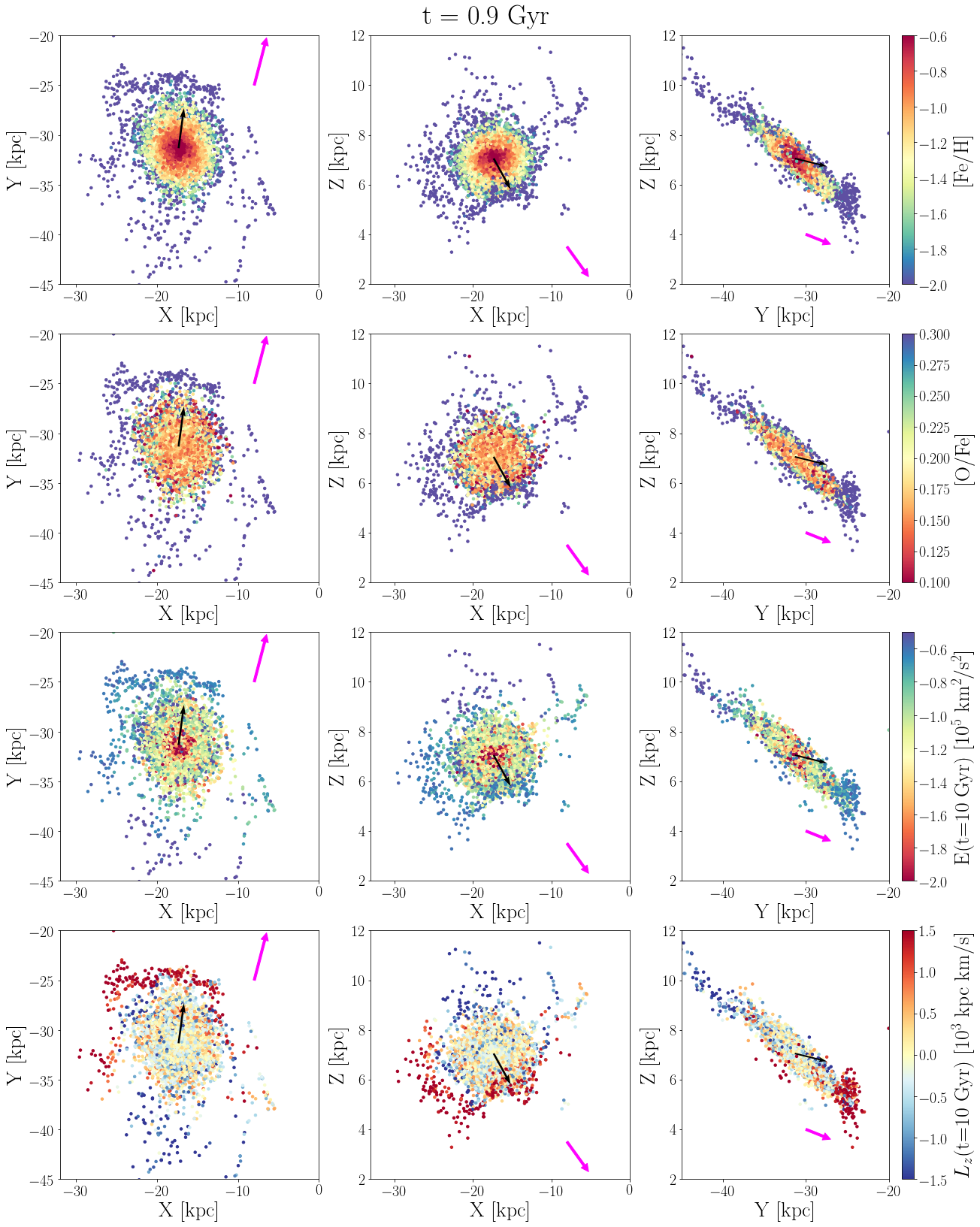}
    
    \caption{Zoom in on the satellite of \fbtdt\, prior to its first pericenteric passage, at $t=0.9$ Gyr. From top to bottom, rows are color-coded based on the stars [Fe/H], [O/Fe], $E(t=10\, {\rm Gyr})$ and $L_z(t=10\, {\rm Gyr})$. The accreted stars that end up less bound to the host galaxy are those located on the outskirts of the dwarf. The dwarf center-of-mass velocity vector is shown as the black arrow. The magenta arrow points towards the host galaxy's center. An animation of the models evolution can be seen \href{https://amarante.netlify.app/GASTRO}{here}. }
    \label{fig:zoom}
    \end{figure*}
All the models have a significant radial metallicity gradient where those with low feedback have steeper gradients compared to their high feedback counterparts. This happens because stronger feedback suppresses star formation more efficiently, and thus delays the enrichment of the central parts of the dwarf, resulting in shallower metallicity gradients. We also measured the metallicity gradient just before the time of the merger and, as the satellite becomes more pressure supported (see Section \ref{sec:dwprofile}), its gradient approaches $\approx -0.04$ dex/kpc in all models, similar to what \citet{naidu+2021_nbody} estimated for GSE. Finally, we also verified that, when run in isolation up to $t=10$ Gyr, the dwarf model's radial metallicity profile becomes shallow, $\approx -0.03$ dex/kpc, in agreement with observations of local dwarfs \citep[e.g.][]{kirby+2013}. \par
Similarly, we fit a linear function to the [O/Fe] radial abundance profile. The [O/Fe] radial gradient is shown in Table \ref{tab:models}. All the models exhibit a very mild gradient consistent with a flat [O/Fe] radial profile. We note, however, that the stars in the outskirts of the dwarfs have higher [O/Fe] in comparison to its central parts. This is due to the lower efficiency of star formation at larger radii which
inhibits star formation, leading to a slow chemical enrichment. We illustrate the chemical profiles of a dwarf in the first two rows of Figure \ref{fig:zoom}. It shows a zoom-in of the satellite in \fbtdt\, -- all the dwarf models show similar trends -- where each star-particle is colored by its [Fe/H] and [O/Fe] (first and second rows respectively).  \par
We have described the structure and chemical evolution of the dwarfs up until just before they merge. In the following we describe the accreted stars after the merger.
\subsection{The accreted stars}
Throughout the paper, we refer to ``accreted stars" as those bound to the host galaxy potential at $t=10$ Gyr originating from the satellite. As the dwarfs in the models evolve, they not only have stars stripped by the host galaxy, but also form new stars after the first pericenter passage. Therefore, we flag as accreted stars those present in the dwarf immediately before its first pericenter passage, and also those born afterwards, up until the satellite disruption. \par
\begin{figure*}
    \includegraphics[width=\textwidth]{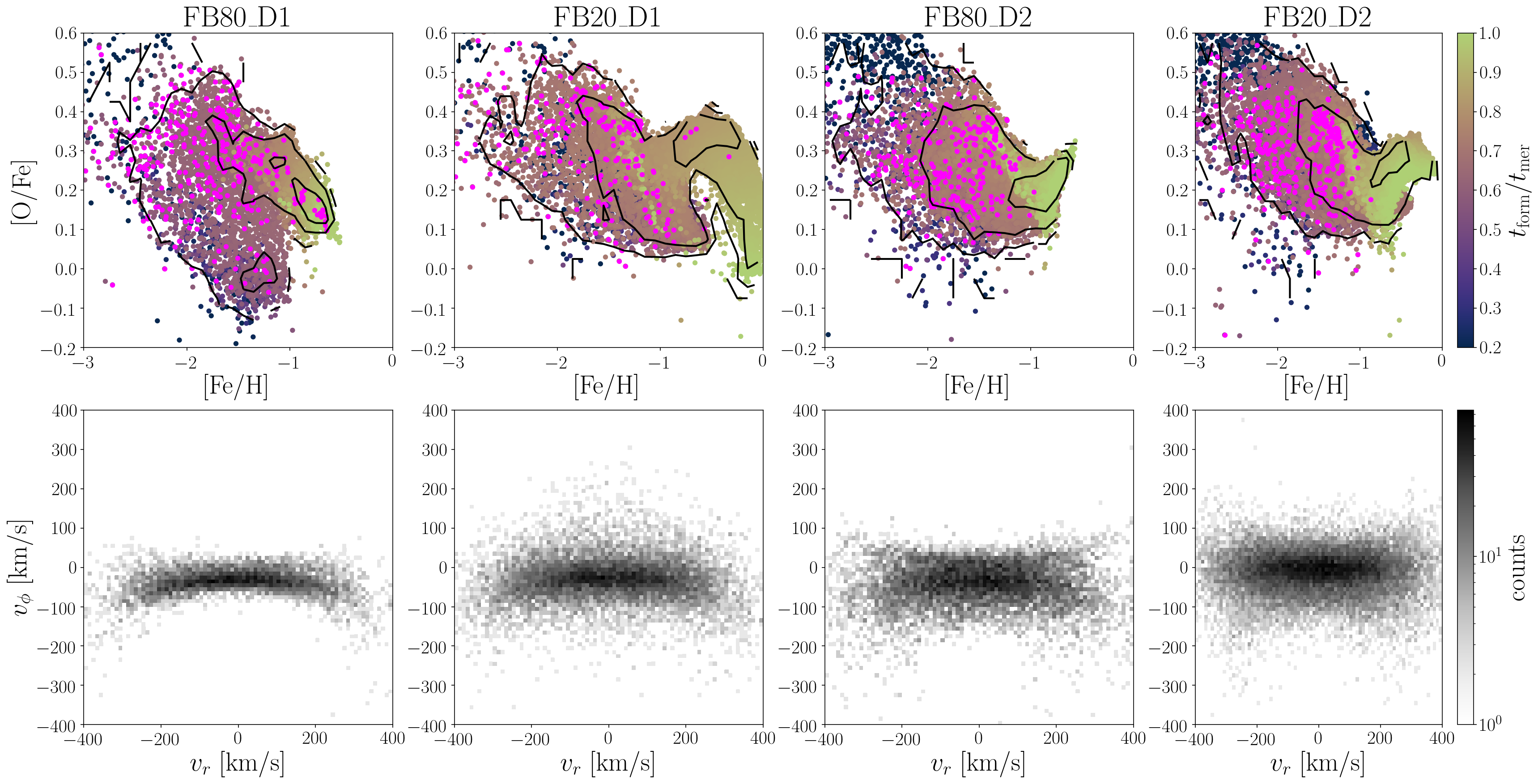}
    \caption{\textbf{Top row:} [Fe/H]-[O/Fe] plane for the accreted stars at $t=10$ Gyr color-coded by the time of formation normalised by the dwarf's time of merger. The non-trivial chemical patterns of the dwarf evolution, e.g. the secondary younger sequence in the chemical abundance, are due to stars formed immediately before the second pericenter. The magenta points are the accreted stars with $L_z < -1.5 \times 10^3$ \lzuni\, and spread over the same [Fe/H]-[O/Fe] space similar to what is observed in the MW stellar halo, see \ref{sec:sequoia} for discussion. \textbf{Bottom row:} $v_r-v_{\phi}$ plane for the accreted stars at $t=10$ Gyr. All the models have the generic ``sausage'' shape, characterized by a radially anisotropic velocity distribution, similar to the GSE merger in the MW. The larger radial velocity dispersion compared to the rotational velocity dispersion is the result of the merging dwarf's radial orbit. The lower feedback models (second and fourth columns) present a larger spread in $v_{\phi}$ compared to higher feedback models (first and third columns). }
        \label{fig:fehofeplane} 
    \end{figure*}

We start by analyzing their chemical abundances and in Table \ref{tab:models} we give the median values of [Fe/H] and [O/Fe] for all of our models. We also show, in the top row of Figure \ref{fig:fehofeplane}, the [O/Fe]-[Fe/H] plane for the accreted stars. Each star is colored by its formation time, $t_{\rm form}$, normalized by the time of the satellite's merger. The higher density region, which is detached from the main chemical sequence, has the youngest stars in the dwarf, which are formed in the central starburst after the first pericenter passage \citep[for details of this process, see][]{du+2019}. This effect is more intense in the low feedback regime, where the stars can reach higher metallicities. Although these seem a rather unusual effect, we note the Large Magellanic Cloud also presents a ``bump" in the [$\alpha$/Fe]-[Fe/H] plane which is associated with a burst in its star formation history \citep{nidever+2020}, and is also observed in MW in-situ stars \citep[prograde stars with $ecc <0.8$, as seen in][]{conroy+2022}. \par
The models presented here are idealized to mimic the GSE merger event in the MW, in terms of the dwarf's mass, orbit and time of merger. As expected, they reproduce the large radial anisotropy in the velocity distribution as observed in the MW. The bottom row of Figure \ref{fig:fehofeplane} shows the distribution of the accreted stars in the $v_r - v_{\phi}$ plane. The radial orbit of the satellite prior to the merger and its relatively high stellar mass causes the final velocity distribution of the accreted stars to be very radial \citep[as demonstrated by][with $N$-body only models]{amorisco2017}. In fact, the ``Sausage"-shape of the velocities distributions was a smoking gun to confirm that the inner stellar halo consists of stars mainly from a single massive progenitor \citep{belo2018}, as several minor mergers are expected to fall with different radial directions assembling a more isotropic stellar halo.  \par
The accreted component in \fbtdt\, has a mean net zero rotational velocity, similarly to what is measured for the GSE \citep[e.g.][]{belo2018, helmi2018, buder+2022}, but model \fbedt, which has the same satellite orbit, shows a slight retrograde net motion. This is also the case for models with dwarf \done, with the stronger feedback leading to the most retrograde realisation of the GSE. As shown in Section \ref{sec:dwprofile}, the lower feedback regime has a higher stellar density in its center allowing it to be more efficiently radialized (also seen in the bottom row of Figure \ref{fig:lzevol}). 
\section{Present-day chemodynamical properties}\label{sec:results}
We now explore the main chemodynamical trends of each model's host galaxy stellar halo, i.e. the accreted stars, after 10 Gyr of evolution. We calculate the gravitational potential and actions at 10 Gyr -- assuming an axisymmetric system -- and integrate orbits of star-particles using {\tt AGAMA} \citep{agama}. \par
In our coordinate system, the vertical component of the angular momentum, $L_z$, represents prograde (retrograde) motion if $L_z>0$ kpc km/s ($L_z<0$ kpc km/s). Finally, throughout this section we also show a sample of chemically selected accreted stars from APOGEE DR17 \citep{apogeeDR17} -- we describe the data selection in Appendix \ref{appen:apogee} -- to illustrate the general properties of the MW stellar halo in the chemodynamical spaces we explore with the models. %
\subsection{Energy-$L_z$ space}\label{sec:e-lz} 
Energy, $E$, and \lz\, are integrals of motion in a time-independent axisymmetric potential. Thus, under the assumption that the host galaxy is not strongly perturbed after the merger, and as long as non-axysimmetric features such as the bar and spiral arms can be neglected, $E$ and \lz\ are approximately conserved. For this reason it is commonly said that the observed dynamic state of stars retains ``memory of their origin" \citep[e.g.][]{helmi2020}. \par
As each star will retain information about its origin, overdensities in the $E-L_z$ plane may be associated to substructures with, in principle, distinct origins. In fact, this plane is commonly used to find substructures in the MW stellar halo \citep[e.g.][]{helmi-zeeuw2000,koppelman+2018,horta+2021}. However, a single massive merger can spread over a large range of $E$ and $L_z$ \citep[see e.g.][]{jean-baptiste+2017,naidu+2021_nbody}. As we will see below, this is clearly seen in our $N$-body + SPH GSE-like merger models which also show similar chemical behavior. \par
\begin{figure*}
    \includegraphics[width=\textwidth]{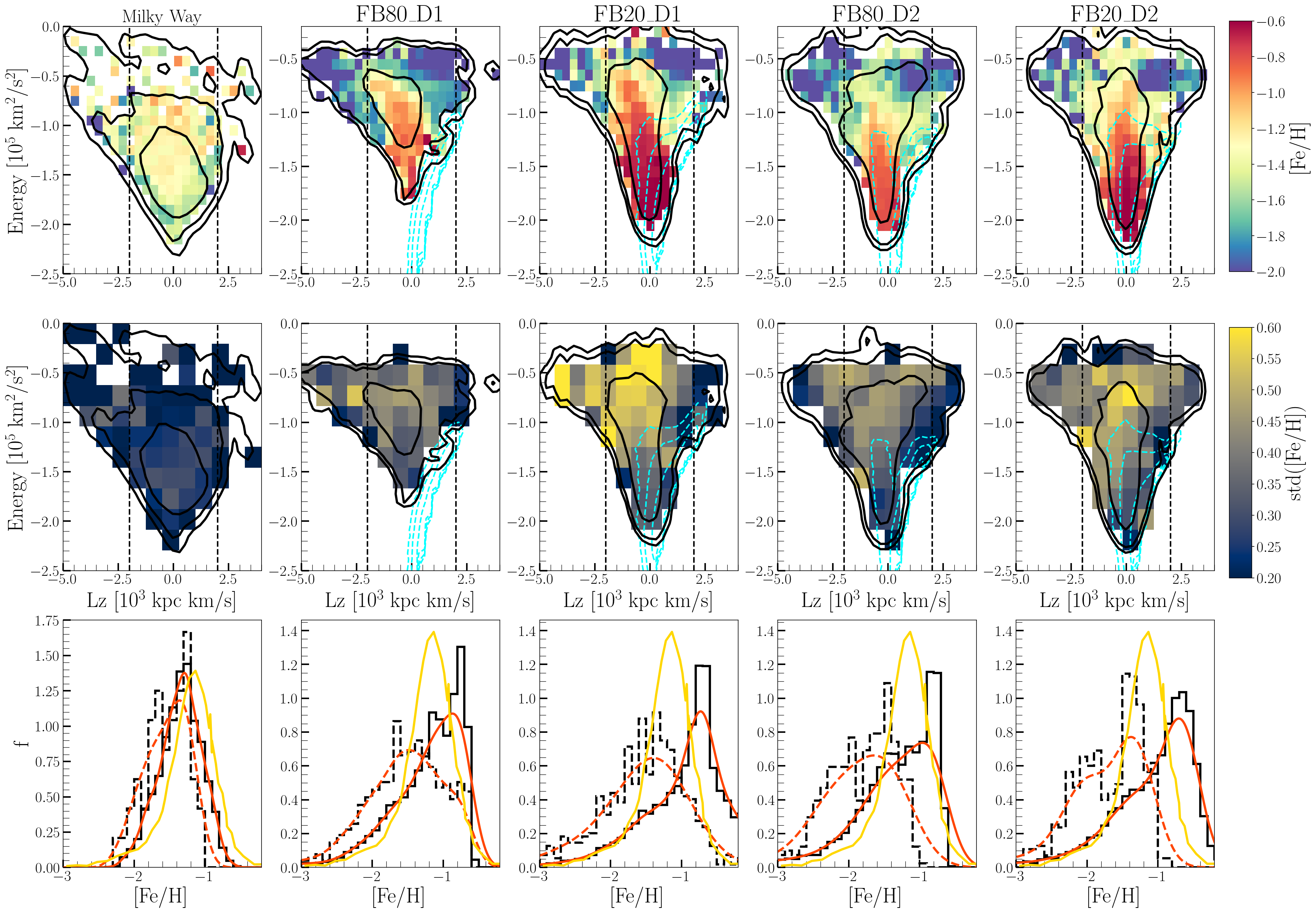}
    \caption{\textbf{First row:} properties of accreted stars in $L_z - E$ space for each of the models at $t=10$ Gyr. The log-density contours of the accreted stars and old stellar disk are shown in black and cyan lines, respectively. Each pixel on the plane is colored by the median [Fe/H] of the accreted stars. It shows how a single merger can spread in this plane with distinct chemical trends. \textbf{Second row:} Same as the first row, but now colored by the standard deviation of [Fe/H] in each pixel. \textbf{Third row:} the [Fe/H] distribution  of the accreted stars and its smoothed KDE version are shown as solid black and orange lines, respectively. For comparison, the MW observational data from the H3 survey \citep{naidu+2020} is shown as the yellow line. The dashed black and orange lines show the distribution and its smoothed KDE version for accreted stars with very retrograde orbits, $L_z <-1.5\times 10^3$ \lzuni. See text for details. }
    \label{fig:enlz}
    \end{figure*}

The top panels of Figure \ref{fig:enlz} show the $E-L_z$ plane of the accreted stars in the MW and in each of our models, at $t=10$ Gyr, where each pixel is color-coded by the median metallicity. We opt to use the same [Fe/H] colorbar scale for the MW and the models for a better direct comparison, although the models have a higher median metallicity compared to the MW, as we will discuss below. Black and cyan lines represent contours of the number-density distribution for all the accreted stars and old disk (stars born in the host galaxy within the first 3 Gyr), respectively. All models have a main overdensity of accreted stars concentrated at $L_z \approx 0$ \lzuni\, and $E > -1.5\times 10^5$ ($-2\times 10^5$) \euni\, in the high (low) feedback models. Despite all the dwarfs having the same initial orbit circularity, $\eta=0.3$, in Figure \ref{fig:appen:enlz} we see that this overdensity is also present for a more circular initial orbit, $\eta=0.5$. The most metal-rich stars are concentrated at $L_z \approx 0$ \lzuni, and are more bound to the host galaxy's potential. On the other hand, the metal-poor stars are less bound to the potential and are often located on very retrograde orbits, $L_z < -2 \times 10^3$ \lzuni, but some also have prograde motion, $L_z>1 \times 10^3$ \lzuni. In the MW, the accreted star candidates in APOGEE also produce a main overdensity at $L_z \approx 0$ \lzuni, but show a trend where the more bound stars are more metal-poor. These are stars associated with ``Heracles" which we discuss in Section \ref{sec:heracles}. The retrograde region, $L_z < -1.5 \times 10^3$~\lzuni\, associated with Sequoia (see Section \ref{sec:sequoia}) is overall at the more metal-poor regime.\par
A model's chemical trend is caused by the radial metallicity gradient present in the dwarf prior to the merger, as discussed in Section \ref{sec:dwchem}. We illustrate this behavior by returning to the zoom in of the dwarf in model \fbtdt\, prior to its merger. The last two rows of Figure \ref{fig:zoom} show each star of the dwarf colored by its $E$ and $L_z$ at $t=10$ Gyr. As discussed in Section \ref{sec:evolution}, the most metal-rich stars are mainly located near the dwarf's center and because they are more bound to the satellite's potential, they are less easily stripped. These stars only become unbound in the very last stages of the merger, ending up being the most bound at $E(t=10)\approx -1.7 \times 10^5$ \euni, with $\langle L_z(t=10) \rangle \approx 0$ \lzuni.\par
On the other hand, the outskirts of the dwarf are populated by metal-poor stars and these get stripped during the first and second pericenter passages, which is why they end up being less bound to the galactic potential at $E>-1.2\times 10^5$ \euni, and with $|L_z| > 2\times 10^3$ \lzuni.\par
The features shown in the top row of Figure \ref{fig:enlz} demonstrate that a single massive merger does not have a unique ``signature" in the chemodynamical space. For instance, all the models present a very retrograde region, $L_z < -1.5\times 10^3$ \lzuni\, and $E>-1\times 10^5$ \euni, with median [Fe/H]$\lesssim -1.8$ that could naively be misinterpreted as a distinct merger event from the main overdensity at $L_z\approx 0$ \lzuni\, due to their distinct metallicities and orbital characteristics (see discussion in Section \ref{sec:sequoia}).\par
We can also see the effects of different feedback strengths. First, high feedback prevents the dwarf galaxy from sinking deeper into the potential compared with the low feedback case. This happens because the high feedback models have less dense centers, due to the reduced star formation (see Section \ref{sec:dwprofile}), and thus are disrupted at larger radii. Secondly, and for similar reasons, low feedback allows the dwarf (and the progenitor) to have a faster metallicity enrichment. The faster increase of metallicity in the low feedback regime also happens more intensively in the central parts of the dwarf (see Section \ref{sec:dwprofile}) which produces a lower dispersion in [Fe/H] for the most bound accreted stars in these models, as shown in the second row of Figure \ref{fig:enlz}. The difference in the efficiency of star formation is also seen when comparing the same model with different feedback strength. The lower feedback strength regime ({FB20}) has an overall higher [Fe/H] dispersion compared to the higher counter-part ({FB80}). This is also evident in the high $|L_z|$ regions, although with a milder difference between the two feedback strengths. For instance, the [Fe/H] dispersion in the prograde/retrograde regions of \fbtdo\, (\fbtdt) is higher by a factor of 0.03 (0.04) dex compared to \fbedo\, (\fbedt). This is caused by the lower gas fraction in the satellite's outer regions (see Figure \ref{fig:dwarfprofile}) where the majority of the high $|L_z|$ stars originate from, as discussed previously in this section.
Lastly, the old disk is hotter in the low feedback strength regime, i.e. it extends to lower $L_z$ and higher $E$ compared to the high feedback strength models, due to the presence of clumps in the progenitor during the first Gyr of evolution.\par
The bottom panels of Figure \ref{fig:enlz} show the metallicity distribution function (MDF) of the accreted stars as solid histograms. The effect of feedback on the MDF of the dwarfs is again seen in the presence of an extended tail towards higher [Fe/H] in the MDF of the low feedback strength. These metal-rich stars are formed during the starburst at the final pericenter. Moreover, although the peaks of the MDFs in both feedback regimes are roughly similar, the low feedback models have higher $\langle{\rm [Fe/H]}\rangle$ compared to their high feedback counterpart (see Table \ref{tab:models}). We also show the [Fe/H] distribution of the H3 survey \citep{naidu+2020} as the yellow line. Clearly, none of the models reproduces the exact MDF peak location for accreted stars in the MW, however we note that the peak metallicity value is degenerate with the merger timescale. We stress that the goal of this work is not to find the exact match between the peak metallicity of a dwarf model and the GSE's, especially given the degeneracy between the time of the merger and the MDF. \par
Nonetheless, we can qualitatively explore trends in the models that are also in the MW, such as the MDF of the very retrograde stars, at $L_z<-1.5\times 10^3$ \lzuni. The H3 survey claims three distinct peaks for these stars which were associated with distinct substructures: Arjuna, Sequoia, I'itoi \citep{naidu+2020}. We show the MDF for the very retrograde stars in our models as the dashed histograms in the bottom panels of Figure \ref{fig:enlz}. The very retrograde stars in all the models are dominated by the metal-poor accreted stars. In the bottom panel of Figure \ref{fig:zoom}, we can see that these stars originated from the more metal-poor outer disk of the satellite, and were the first to be stripped, retaining some of their orbital angular momentum. We note that it is also tempting to visually identify ``peaks" in the MDFs, especially for \fbtdt. However, if we either increase the bin size\footnote{The bin size is 0.1 dex as in \citet{naidu+2020}.}, or smooth the distribution with a kernel-density estimator (KDE, orange lines in the bottom panels of Figure \ref{fig:enlz}), with the bandwidths chosen using Scott's rule \citep{scott2005}, the MDF shows a smoother behavior with a single-peak and extended tail towards low [Fe/H] distribution. We also note that the MDF for the retrograde stars in APOGEE may also hint at visually identified peaks, but the smoothed KDE MDF simply demonstrates that these have more metal-poor stars compared to the whole distribution. We will discuss whether there is any statistical significance of these ``peaks" in Section \ref{sec:sequoia}.
\subsection{Eccentricity distribution}\label{sec:ecc} 
We calculate the eccentricity as $ecc = (r_{apo}-r_{per})/(r_{apo}+r_{per})$, where $r_{peri}$ and $r_{apo}$ are the pericenter and apocenter in galactocentric spherical coordinates of the stellar orbit.
In the MW GSE stars are found on very eccentric orbits, with $ecc>0.7$, which is a consequence of the dwarf's initial radial orbit. For this reason, the eccentricity of stellar orbits is commonly used to select samples of GSE stars \citep[e.g.][]{mackereth+2019,mackereth_bovy_2020,naidu+2020}. Besides that, Sequoia, which is in the very retrograde part of the accreted halo \citep[e.g.][]{myeong+2018_sequoia}, also has less eccentric stellar orbits when compared with GSE. In this section we explore the chemical trend with eccentricity of the accreted stars.\par
\begin{figure*}
    \includegraphics[width=\textwidth]{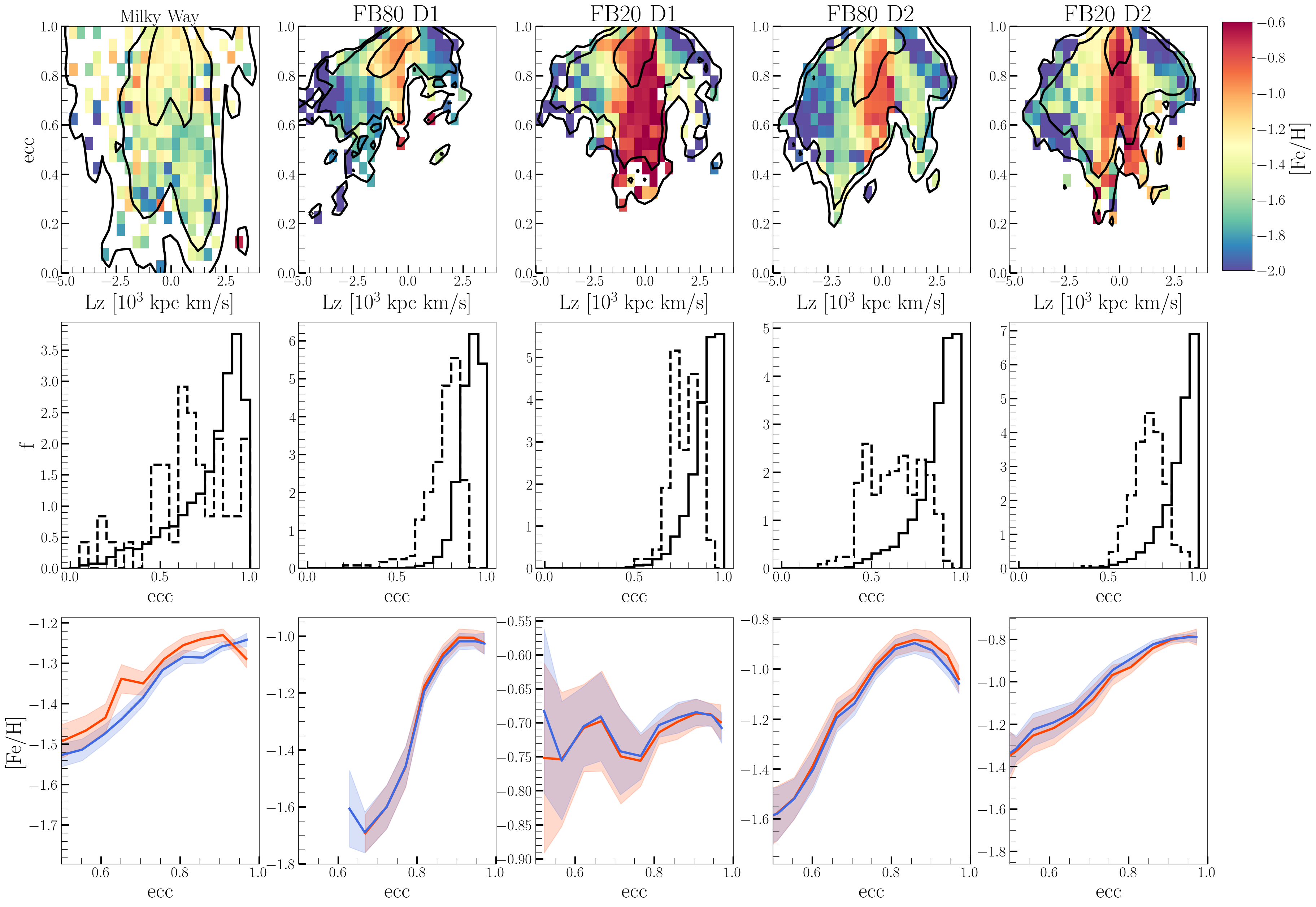}
    
    \caption{\textbf{First row:} properties of accreted stars in $L_z-ecc$ space for each of our models at $t=10$ Gyr. The log-density contours of the accreted stars are shown as black solid lines. Each pixel on the plane is colored by the median [Fe/H] of the accreted stars. \textbf{Second row:} the accreted stars $ecc$ density distribution. Solid and dashed lines show the distribution for all the stars and those with $L_z <-1.5\times 10^3$ \lzuni, respectively. While the bulk of stars have $ecc>0.7$, the orbit for very retrograde stars are less eccentric, similar to what is observed in the MW. \textbf{Third row:} $ecc-$[Fe-H] relation for all the accreted stars and its retrograde ($L_z<0$ \lzuni) component within the Solar neighborhood for each model shown in blue and red respectively. The shaded area corresponds to $\sigma/\sqrt{N}$, where $\sigma$ and $N$ are the [Fe/H] dispersion and total number of stars in each bin, respectively.}
    \label{fig:ecc}
    \end{figure*}

The first row of Figure \ref{fig:ecc} shows the $L_z-ecc$ plane  where each pixel is colored by the median [Fe/H] of the accreted stars in the MW and in the models (at $t=10$ Gyr). The number-density distribution is shown as black solid lines. The chemically selected MW accreted star candidates have a main overdensity at $ecc>0.7$ and a large tail towards lower eccentricities, possibly associated with other accretion events or with an ancient in-situ population formed prior to the MW disk \citep{belokurov+2022,myeong+2022}. It is also evident that the lower eccentricity stars, $ecc<0.7$, are more metal-poor compared to the higher eccentricity ones. In the models, the bulk of the distribution is centered at $L_z\approx 0$ \lzuni\, and $ecc>0.8$, which results from the radialization of the satellite's orbit during the merger process (see Section \ref{sec:evolution}). Nonetheless there is still a small, but non-negligible fraction of stars with $ecc<0.7$.  \par 
The second row of Figure \ref{fig:ecc} shows the $ecc$ distribution for all the accreted stars and those with $L_z <-1.5\times 10^3$ \lzuni, as solid and dashed black lines respectively. While the accreted stars have median $ecc=0.9$ in all the models, those with $|L_z| > 1.5\times 10^3$ \lzuni\, tend to be on less eccentric orbits, with median $ecc=0.7$, but with a significant tail towards $ecc\approx 0.5$, despite being less bound to the galactic potential notice, in the top row of Figure \ref{fig:enlz}, that $E$ tends to higher values for $|L_z|>1.5 \times 10^3$ \lzuni\,). This reflects the fact that the stars in the outskirts of the dwarf are more rotationally supported, having \vrotsig$>2$, and retain more of their angular momentum.\par
The third row of Figure \ref{fig:ecc} shows, for the accreted stars, the relation between the median [Fe/H] and their orbital eccentricity. In the MW, which is shown in the first column, [Fe/H] increases with  stellar eccentricity (blue line). \citet{kordopatis2020} have already shown this positive gradient for kinematically selected ($v_{\phi}<0$ km/s) accreted star candidates. We further verify that the retrograde accreted stars in our chemically selected sample also have a positive gradient in the [Fe/H]-$ecc$ relation, as shown by the red line.\par
For the models, we select the accreted stars within a mock Solar neighbourhood, i.e. $5 <R/{\rm kpc}<12$ and $|z|<3$ kpc, and show the relation for all the accreted stars and the retrograde ones, indicated by the blue and red lines respectively. The shaded area represents the regions between the 16\% and 84\% percentiles of the [Fe/H] for a given eccentricity bin. Remarkably, for a given dwarf with the same initial orbital conditions, the feedback changes the [Fe/H]-ecc relation significantly. Both high feedback models show the same general trend: an increase of [Fe/H] with $ecc$, until it appears to turn over at $ecc\approx 0.95$.  On the other hand, the low feedback models present different trends: \fbtdo, which ends the merger later, has a flat $ecc$-[Fe/H] relation, whereas \fbtdt\, shows a positive trend over all the $ecc$ range. \par
The difference in the $ecc-$[Fe/H] relation between different feedback models is again caused by the higher star formation rate in the low feedback regime. As shown in Section \ref{sec:dwprofile}, after the first pericenter the dwarf in the low feedback models has a higher star formation rate compared to their high feedback counterpart. This process forms more metal-rich stars in the dwarf's center and, as shown in Figure \ref{fig:zoom}, these stars remain bound longer in the satellite and end up being the most radialized, thus having higher $ecc$. However, the very eccentric stars, $ecc\gtrsim0.95$, are overall formed earlier (i.e are older) compared to those with $ecc<0.95$, and were more bound in the dwarf, causing the turnover in the $ecc-$[Fe/H] relation. This is enhanced in the high feedback regime due to the more inefficient star formation, compared to the low feedback, and the stars with $ecc>0.95$ never reach higher [Fe/H] abundances.
\subsection{Action Space}\label{sec:action} 
We now focus on the distribution of the merger debris in action space. In general, orbits in a nearly-integrable gravitational potential can be described by three actions which are adiabatic invariants in a slowly varying potential \citep{bt2008}. The axisymmetry of the interpolated potential, as described in the beginning of this Section, allows us to calculate the actions assuming the local St{\"a}ckel Fudge approximation \citep{binney2012}. In such a gravitational potential, the action $J_R$ measures the cylindrical radial excursion of an orbit, $J_{\phi}$ quantifies the circular motion and is equal to $L_z$, and $J_z$ measures the vertical excursion of the orbit with respect to the galactic plane. These have been used to study the stars in the MW stellar halo \citep[e.g.][]{posti+2018,feuillet+2020,feuillet+2021, lane+2021}. Thus in this section we explore the action space of the accreted stars in our models. \par  
\begin{figure*}
    
    \includegraphics[width=\textwidth]{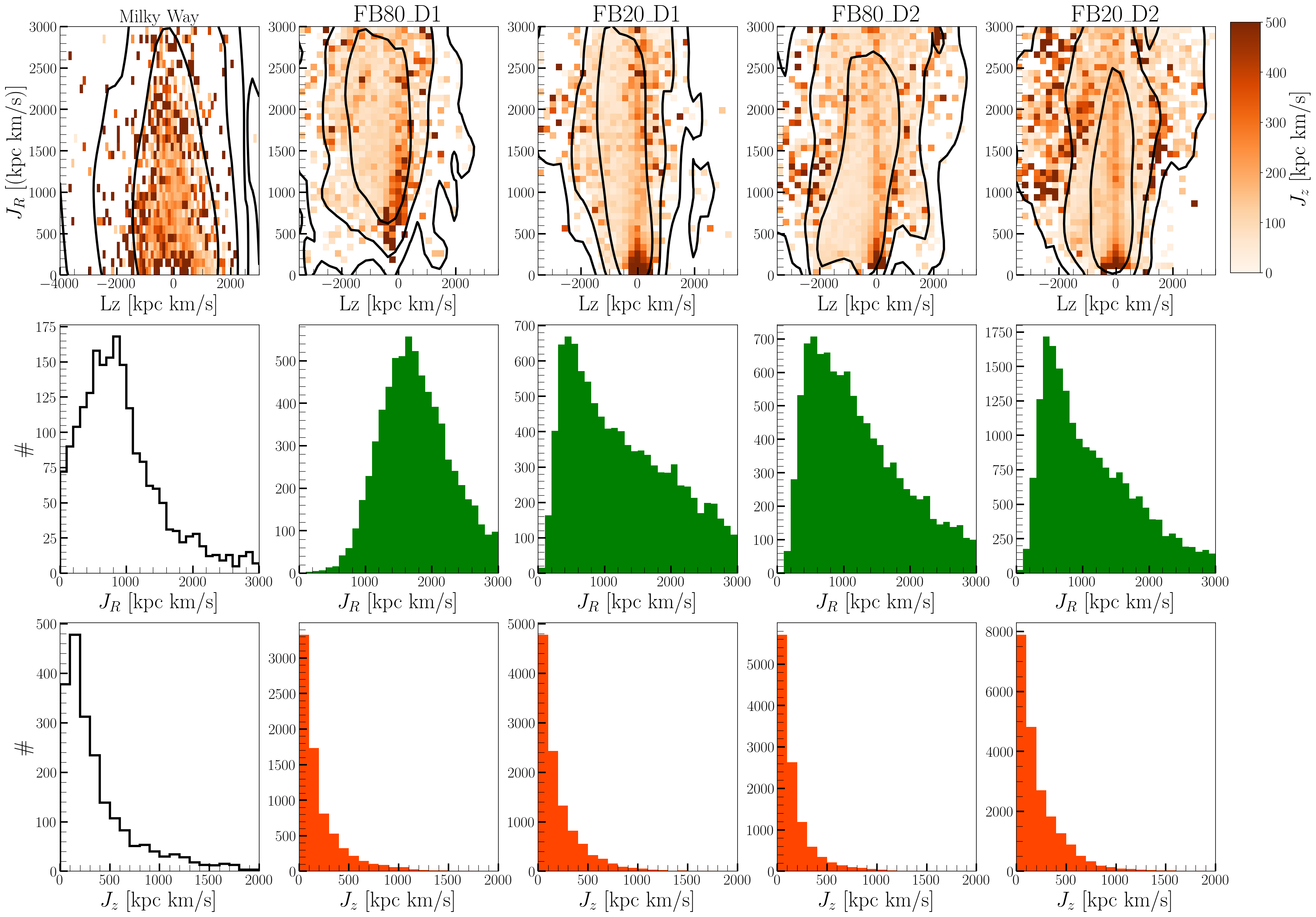}
    
    \caption{First row shows the $L_z - J_R$ plane for the accreted stars at $t=10$ Gyr. Each star is colored by its vertical action, $J_z$. Stars with net zero rotation spread over a large range of $J_R$ and $J_z$, whereas the very retrograde ones, $L_z < 2 \times 10^3$ \lzuni\, tend to have higher $J_z$ component. We restrict ourselves to show stars with $J_R<2000$ \lzuni\, because of the scarcity of stars observed in the MW with high $J_R$ values. The second and third rows show the histogram of $J_R$ and $J_z$ respectively. A GSE-like merger spreads over a large range in $J_R$ while remaining relatively confined in $J_z$. }
    \label{fig:action} 
    \end{figure*}
The top row of Figure \ref{fig:action} shows the $L_z$-$J_R$ plane where each star particle is colored by its vertical action, $J_z$. Given the radial and low inclination nature of the orbit of the merging dwarf, it is expected that most of the accreted stars will tend to have large values of $J_R$ compared to $J_z$. Indeed, the accreted stars spread over a larger range of $J_R$, as seen in the second row of Figure \ref{fig:action}, than what is typically used to define GSE in the MW \citep[$J_R > 900$ \lzuni, e.g.][]{feuillet+2021}. Indeed, our MW chemically selected accreted stars also spread across a large range of $J_{R}$. 

Finally, the majority of accreted stars in our models have relatively low $J_z$, due to the low inclination nature of the merger orbit, as seen in the third row of Figure \ref{fig:action}. This is also observed in the MW accreted sample where the majority of stars have $J_z \lesssim  800$ \lzuni\, in agreement with the models. The MW's tail towards higher $J_z$ is due to the presence of stars originated from distinct merger event(s), such as the Helmi Stream \citep{helmi_streams1999, yuan+2020}. Finally, we note that, in the models, the accreted stars with $|L_z|>2 \times 10^3$ \lzuni\, also have typically low $J_z$, and are found with relatively high $J_R$, despite having on average lower eccentricity (as discussed in Section \ref{sec:ecc}). \par
The fact that a merger like the GSE spreads over such a large range of $J_R$, and as previously shown, in $E$ and $L_z$, reinforces the need to be cautious when trying to find substructures in the MW halo. Nonetheless, the fact it is somewhat restricted in $J_z$ \citep[which is also used as an extra selection criterion for GSE stars e.g.][]{yuan+2020,limberg+2021}, can help interpret the origins of some substructures in the MW, as will be discussed in the next section.
\section{Implications for MW substructures}\label{sec:implication}
 
In the previous section we showed and studied the main chemodynamical properties of the stellar halo of several MW-like galaxies that went through a GSE-like merger during their first Gyrs. Now, we explore the implications of our results for the known substructures observed in the MW's stellar halo. Most importantly, we want to ask whether a single GSE-like merger can explain multiple substructures observed in the MW. 
\subsection{Gaia-Sausage-Enceladus}\label{sec:gse} 
As shown in Section \ref{sec:evolution}, the models presented are intended to reproduce general trends observed for the GSE in the MW. 
While all models show a large radial velocity dispersion compared to the rotational velocity dispersion, we note that models with low feedback have a larger spread in $v_{\phi}$ compared to the high feedback regime (see Figure \ref{fig:fehofeplane}). As shown throughout this study, stronger feedback inhibits star formation and thus there is a deficiency of stars at large radii in these models. The stars on the outskirts of the dwarf are the ones which keep more of their initial $L_z$ and end up with $|L_z| \geq 1\times 10^3$ \lzuni\, (see Section \ref{sec:e-lz}), thus causing a larger spread in the $v_{\phi}$ distribution. \par
The MDF is another important constraint on the progenitor dwarf of GSE. However, observational measurements in the MW have not converged yet, as the GSE peak metallicity varies depending on the survey and the selection criteria for GSE star candidates (as pointed out in Section \ref{sec:intro}). In the models presented here, the final MDF is degenerate with both the time of the merger and the feedback model adopted in the model's subgrid physics. The idealized nature of the models allows us to adjust the orbit of the dwarf, which will lead to an early/late merger time permitting us to fine-tune the MDF peak. \par
The [Fe/H]-$ecc$ space also depends on the adopted feedback model as shown in the third row of Figure \ref{fig:ecc} and discussed in Section \ref{sec:ecc}. In the MW, \citet{kordopatis2020} found a positive correlation for a sample of retrograde GSE candidate stars, similar to what we see in model \fbtdt. The other models either show a positive trend up to $ecc\approx 0.85$ when the median [Fe/H] starts to decrease for higher $ecc$ (\fbedo\, and \fbedt) or a flat correlation (\fbtdo). These are related to feedback strength and the star formation history of the dwarfs (see Section \ref{sec:ecc}). This suggests that the [Fe/H]-$ecc$ correlation is a strong constraint on the initial conditions of the GSE merger and the feedback prescription.
\subsection{Sequoia and other very retrograde substructures}\label{sec:sequoia}
Since the early 2000s, there has been growing evidence that very retrograde stars in the MW stellar halo are more $\alpha$-element depleted compared to the bulk of the halo  \citep{stephens_boesgaard2002, venn+2004}. It was speculated that these stars originated from different accretion events during the MW stellar halo build up. More recently, \citet{myeong+2018_sequoia, myeong+2019} found evidence that the very retrograde high energy (VRHE) stars are chemically distinct from the GSE and suggested they originated from a different progenitor, which they named Sequoia. The fact that Sequoia is more metal-poor compared to the bulk of GSE stars has been confirmed by several other studies \citep[][]{myeong+2019,ruiz-lara+2022}. However, it is still debatable how Sequoia compares to GSE in terms of its $\alpha$-element abundance. While \citet{mackereth+2019}, \citet{myeong+2019}, and \citet{feuillet+2021} all found the retrograde stars associated with Sequoia to be more $\alpha$-rich compared to the bulk of the GSE, \citet{matsuno+2019}, \citet{monty+2020} and \citet{matsuno+2021} found evidence for Sequoia being more $\alpha$-poor compared to GSE. Interestingly, \citet{aguado+2020} using high resolution spectroscopy, found that Sequoia and GSE have rather similar abundance of n-capture elements, although there is a hint for a distinct [Ba/Fe] between them. Recently, \citet{horta+2022} and \citet{limberg+2022}, using APOGEE DR17, showed that Sequoia and GSE are chemically indistinguishable from each other.\par
The [O/Fe]-[Fe/H] plane of the accreted stars in each model is shown in the top row of Figure \ref{fig:fehofeplane}. The VRHE stars in the models, defined as the accreted stars with $L_z < -1.5 \times 10^3$ \lzuni, are shown as the magenta points. Similar to what is observed in the H3 survey (see figure 14 in \citealt{naidu+2020}), they spread over a large range in this plane, and could be misinterpreted as originating from a different progenitor. As discussed in Section \ref{sec:dwchem}, the gradient present in the radial abundance profile of the dwarfs prior to the merger is significant enough to create such intricate chemical patterns in the accreted halo.\par
One such pattern is the observed metallicity of the VRHE stars in the MW. These stars, which are, for instance, associated with the Sequoia, are more metal-poor compared to the GSE and their MDF hints at the presence of three peaks, as observed with the H3 Survey (see figure 14 in \citealt{naidu+2020}). Due to this, \citet{naidu+2020} suggested the presence of two other substructures, Arjuna and I'itoi, besides the previously known Sequoia, each associated with the visually identified peaks. While all our models are consistent with the fact that the VRHE stars are more metal-poor compared to stars with $L_z \approx 0$ \lzuni\, (Section \ref{sec:e-lz}), we further investigate whether any of them have a MDF with a statistically significant multiple-peak structure. \par

To do this, we use the test proposed by \citet{silverman1981}\footnote{The python code for the test is available at\\ \url{https://github.com/lberaldoesilva/silverman-test}}, in which there is no assumption on the form of the data probability distribution function,  as opposed to the often-used Bayesian Information Criterion of a Gaussian Mixture Model (see a summary on the method in \citet{beraldo+2021}). These tests reveal no significant evidence for the presence of multiple peaks in the MDFs of VRHE stars in our models. This serves as a caution on the statistical significance of visually identified peaks. It remains to be seen if the peaks observed in the H3 survey are statistically significant and whether they correlate with the star formation history of the structures (in our case we did not find any such correlation, as discussed in Section \ref{sec:e-lz}).  \par
Another characteristic of Sequoia and other VRHE substructures that has been used to argue for distinct entities is the fact that they have lower eccentricity compared to the GSE \citep[e.g.][]{myeong+2019,limberg+2021}. In section \ref{sec:ecc} we showed that VRHE stars in our models are on less eccentric orbits compared to the bulk of the accreted stars (see Figure \ref{fig:ecc} second row). This is expected as these least bound stars are the first to be stripped, retaining some of their orbital angular momentum and thus end up on less eccentric orbits \citep[as already shown in][with $N$-body only models]{naidu+2021_nbody}. The self-consistent chemistry of our models then demonstrates that a radial [Fe/H] gradient naturally appears in the merging dwarf, giving rise to the metal poor nature of the VRHE stars. \par
The common origin of GSE, Sequoia, and perhaps other VRHE structures, then should result in a relatively small spread in the ages of stars of each substructure. The mean age of GSE stars is found to be $\approx 10$ Gyr \citep[e.g.][]{kilic+2019,borre+2021,montalban+2021, matsuno+2021_ages}. On the other hand, Sequoia stars are found to be slightly older than GSE \citep[e.g.][]{feuillet+2021} and with a hint of being accreted circa 0.3 Gyr earlier than GSE \citep[][but this small difference is within the error of the estimated values]{kruijssen+2020}. Our models can naturally explain the slight difference in age and time of accretion measured in the literature. First, as shown in Section \ref{sec:e-lz}, the VRHE originated from the outskirts of the dwarf and these are the less bound stars and thus are stripped first. Secondly, in Section \ref{sec:dwprofile} we show that, during the merger process, there is a burst of star formation mainly in the central parts of the dwarfs. Consequently these relatively younger stars will end up in the region mainly associated with GSE, i.e. at $|L_z|<0.5$ \lzuni. \par
We can also constrain the relative mass fraction of VRHE substructures in comparison to GSE. \citet{naidu+2020} estimated the relative fraction of VRHE stars in comparison to GSE to be $<2.3\%$, i.e. much smaller than initially estimated \citep[e.g. see][]{kruijssen+2020}. All our models show a relatively small fraction of VRHE compared to the rest of accreted stars, with values ranging from 1.3 to 4.2\% (see Table \ref{tab:models}). We note that we are showing the overall contribution without taking into account possible effects of a survey selection function. \par
We have demonstrated, with a set of \gastro\, library models, that VRHE accreted stars share the same qualitative properties as those observed in the MW: they are more [Fe/H]-poor and [$\alpha$/Fe]-rich compared to the bulk of the accreted stars, but spread over a large range in the abundance plane (top row in Figure \ref{fig:fehofeplane}), have an apparent multi-peak MDF, are on less eccentric orbits and comprise similar mass fraction compared to GSE. We also note that the large spread in the chemical space is expected for relatively large dwarfs \citep[see, e.g.][review]{tolstoy+2009}. Thus, our results suggest that Sequoia, Arjuna, I'itoi, and other retrograde structures with these characteristics \citep[e.g.][]{oria+2022,dodd+2022} are naturally explained as stars originating from the outskirts of the GSE satellite. This was shown in \citet{koppelman+2020} and \citet{naidu+2021_nbody} with $N$-body only models; here we have demonstrated the same result with models where the satellite (and the disk) has a self-consistent metal enrichment. 
\subsection{Heracles}\label{sec:heracles}
While most of the substructures are mainly observed at $R\geq5$ kpc, the inner parts of our Galaxy may also hide another significant merger event which happened in the first Gyrs of the MW. Evidence for such a merger was first seen as a low-energy group of globular clusters not associated with any known accreted substructure \citep{kruijssen+2019,massari+2019, forbes2020}. \par
\citet{horta+2021}, using APOGEE DR16 and \textit{Gaia} DR2, found stronger evidence in the chemodynamical space of a merger remnant located in the inner part, $R<5$ kpc, of the Galaxy. They named this inner Galaxy substructure \textit{Heracles} and estimated its progenitor had a stellar mass of $\sim 5\times10^8$ \msun (see also \citealt{kruijssen+2020}), i.e. as massive as the GSE merger event. 
More recently, \citet{lane+2021} showed that the APOGEE survey selection function causes gaps in the $E-L_z$ space. Given the Galactocentric radii distribution of a stellar population, they will be constrained to a parabola in the $E-L_z$ space, where stars at lower radii can reach lower $E$ (see their figure 6). This explains the presence of two groups, low and high energy, in the $E-L_z$ plane of chemically defined accreted stars as observed in \citet{horta+2021}. Despite this dynamical argument of Heracles being the low energy counterpart of GSE, one can still argue about the distinct chemistry of both structures. In the Milky Way, the low energy accreted stars, associated with Heracles, are on average more [Fe/H]-poor and [$\alpha$/Fe]-rich compared to the high energy accreted stars, associated with GSE \citep{horta+2021,naidu+2021}. Moreover, Heracles seems to have a flatter [$\alpha$/Fe]-[Fe/H] relation compared to GSE stars. \par
In Section \ref{sec:e-lz} we showed that the models' metal-rich accreted stars are located preferentially in the low energy part of the $E-L_z$ plane, i.e. the \textit{opposite} of what is observed for Heracles. This result is independent of the dwarf and/or the merger impact parameter explored in the current paper. Therefore our results do not explain the [Fe/H]-Energy relation observed for Heracles in the MW if it is associated with the GSE, as argued in \citet{lane+2021}. \par
Given that the MW's accreted stars are clearly distinct in the [Mg/Mn]-[Al/Fe] plane from in-situ stars \citep[e.g.][]{das+2020, horta+2021, perottoni+2021, naidu+2021} -- this chemical selection of accreted stars also holds for the bulge region \citep{lucey+2022} -- we can conclude that a single merger event such as the GSE is not able to explain all the chemodynamical trends observed in the inner MW. Then, the origin of Heracles can be attributed to a distinct merger \citep[as argued by, e.g. ][]{kruijssen+2019, horta+2021}, or as part of the in-situ turbulent stellar system located in the inner part of the Galaxy \citep{belokurov+2022, myeong+2022}. If Heracles was indeed accreted, such a merger event would imprint an increase in the star formation rate at $R < 5$ kpc \citep{orkney+2022}.
\subsection{Other substructures: a cautionary approach} 
There are several automated techniques used to search for substructures in the MW stellar halo \citep[e.g.][]{koppelman+2019, yuan+2020,limberg+2021,malhan+2022, lovdal+2022}, which are important to elucidate the accretion history of our Galaxy. However, several substructures identified in the MW have a separation in energy, angular momentum and [Fe/H] consistent with the single merger event models presented here. This calls into question the true nature of these substructures as distinct galactic entities. \par
We can use our fine-tuned GSE-like merger models to reconcile substructures into a single merger event. For instance, \citet{malhan+2022} identified Pontus as a new substructure in the dynamical space of a sample of globular clusters \citep{vasiliev_baumgardt2021}. \citet{limberg+2021} also identified substructures, DTG-7, DTG-14, and DTG-18, in a sample of metal-poor halo stars with similar dynamical properties as Pontus. Despite being classified as distinct groups, \citet{limberg+2021} raised the possibility that these were part of GSE given their their low $J_z$ and high orbital eccentricities. \par
Our results also suggest that Pontus, and consequently the other DTG groups, have the same origin as the GSE. For instance, both have $L_z \approx 0$, \lzuni, $J_z < 500 $ \lzuni\, and their difference in $J_R$ is lower than $\sim 1000$ \lzuni. These are within the expectations of a GSE-like merger as shown in Figure \ref{fig:action} and reinforced by the fact that they also have similar chemical abundance \citep{malhan+2022, malhan2022}. Moreover, we verified in APOGEE DR17 that Pontus and GSE display a rather similar MDF, despite the latter having higher mean [Fe/H] \citep[as shown in][]{malhan2022}. It remains to be confirmed with more data whether the [Fe/H] difference is significant enough to rule out a common origin between Pontus and GSE, as suggested by our models. \par
Using the same line of reasoning as above, we can use our models to identify substructures which are \textit{not} associated to the GSE merger event. For example, besides Heracles (see Section \ref{sec:heracles}), we argue that LMS-1/Wukong \citep{yuan+2020,naidu+2020} and Cetus \citep{newberg+2009,yuan+2019,yuan+2022} also most likely have distinct progenitors other than the GSE. \citet{malhan+2022} showed that both substructures have $J_z\approx 2000$ \lzuni, meaning they have large vertical excursions. Throughout this paper we have shown that a single GSE-like merger spreads over a large range in Energy, $L_z$ and $J_R$, but in Figure \ref{fig:action} we show that the accreted stars will end up typically with $J_z<1000$ \lzuni, i.e. it cannot produce structures like LMS-1/Wukong and Cetus. Despite not being associated with GSE, we can question whether they do have a distinct progenitor, based on the similar arguments used for Pontus and GSE. Besides having similar $J_z$, the distance between LMS-1/Wukong and Cetus in $L_z$ and $J_R$ is less than $\sim 600$ \lzuni\, and $\sim 300$ \lzuni. This difference is well within the range of what a merger can produce (as shown in Figure \ref{fig:action}). This argument is reinforced by the fact that they also have only a difference of 0.04 dex in [Fe/H] \citep[see figure 7 in][]{malhan+2022}. \par
Finally, our models also predict the presence of a \textit{prograde} metal-poor accreted population. This somewhat unexpected population would potentially be hard to distinguish from disk contaminants, such as the Splash \citep[][]{dimatteo2019,belo2020}, however the apocenters of its stars should be relatively high, $r_{apo} \approx 25$ kpc, similar to those associated with its retrograde counterpart associated with Sequoia. In fact, we suggest Cand14/DTG-2 \citep{myeong+2018,yuan+2020} as a candidate for being the prograde GSE remnant as it is on a prograde orbit, with $L_z\approx 1500$ \lzuni, $J_R <1000$ \lzuni, $J_z<300$ \lzuni, $ecc\approx 0.6-0.8$, and is more metal-poor, $\langle {\rm [Fe/H]}\rangle \approx -1.45$, compared to the GSE peak metallicity.
\section{Conclusion and final remarks}\label{sec:conc}
In this work, we have used a subset of the \gastro\, library and explored the chemodynamical features of the stellar halo in models in which a GSE-like galaxy merges with a MW-like galaxy. We list below our main conclusions:

\begin{itemize}
    \item The mass density distribution in the dwarfs is more centrally concentrated in the low feedback models after the first pericenter passage (Figure \ref{fig:dwarfprofile}), allowing them to become more radialized, with $\langle L_z \rangle \approx 0$ \lzuni\, (bottom row of Figure \ref{fig:lzevol}), in contrast to their high feedback counterparts;
    \item We confirm with a self-consistent metal enrichment model the predictions from $N$-body only models (under the assumption of an initial radial [Fe/H] gradient) that the accreted stars on low energy orbits are more [Fe/H]-rich than those on high energy orbits (top row of Figure \ref{fig:enlz}); 
    \item Very retrograde high energy stars are on average more [Fe/H]-poor and [O/Fe]-rich, but with a large spread in the chemical abundance plane, and on less eccentric orbits compared to the overall accreted stars distribution (bottom row of Figure \ref{fig:enlz} and second row of Figure \ref{fig:ecc}), similar to what is observed in the MW. This confirms that Sequoia, Arjuna and I'itoi are likely parts of the GSE merger event given its dynamical (as demonstrated in \citealt{koppelman+2020,naidu+2021_nbody}) and chemical properties;
    \item A GSE-like merger has a broad $J_R$ distribution and a narrow one in $J_z$ (Figure \ref{fig:action}). Therefore, we suggest that Pontus \citep{malhan+2022} is most likely part of the GSE merger given its small spread in energy and action space, in addition to having the same [Fe/H] as GSE stars;
    \item None of our GSE-like merger models can explain the chemistry of Heracles as part of the GSE merger event, as argued by \citet{lane+2021}. This could give further credence to a scenario where an earlier massive merger event occurred in the lifetime of the Galaxy \citep[as in][]{horta+2021,naidu+2021}, but it remains to be confirmed whether it has an accreted origin or is part of an old in-situ population \citep[][]{belokurov+2022, myeong+2022};
    \item With a similar line of reasoning, we also suggest that LMS-1/Wukong \citep{yuan+2020lms1,naidu+2020} and Cetus \citep{malhan+2022} most likely share the same progenitor, but different from GSE, as both have a small spread in the action space and have the same [Fe/H]. The high value of their $J_z$ suggests their progenitor had an orbit with a high inclination angle with respect to the Galactic plane, as opposed to the low $J_z$ values of the accreted stars in the low inclination mergers shown here;
    \item We also predict the presence of prograde substructures, with relatively low $J_z$, $ecc\approx 0.6-0.8$ and which are more metal-poor compared to the GSE that originated from the outskirts of the merging satellite. A potential candidate for the prograde remnant of GSE is Cand14/DTG-2 \citep{myeong+2018,yuan+2020}.
    
\end{itemize}
The self-consistent star formation and chemical enrichment of the models presented here provide insights into the structural evolution of the satellite during the merger and is a step further in complexity compared to pure $N$-body models. This clarifies how a single GSE-like merger is able to produce a rich chemodynamical space and thus helping to reconcile some known substructures to a single merger event. It is evident that a single merger event will not explain all the MW stellar halo substructures, but it serves as a cautionary tale when identifying potentially new substructures. Moreover, hydrodynamical cosmological simulations show that MW-like galaxies have their inner ($r<20$ kpc) stellar halo dominated by roughly 3 massive mergers, and low mass accreted satellites only start to dominate the stellar halo fraction in the halo outskirts \citep{fattahi+2020}. These can set constraints on the probability of finding accreted structures from distinct mergers, and stress the need for a validation of substructure-finding algorithms using MW-like formation models.

\acknowledgments
The authors wish to thank the anonymous referee for useful comments that helped improve this work. 
V.P.D. and L.B.S acknowledge that the discussion with James Binney at IAU 353 held in Shanghai 2019 seeded the idea of project \gastro. J.A. thanks Guilherme Limberg and Hélio Perottoni for general scientific discussions, and Zhen Yuan for insightful discussion on stellar halo substructures. C.L. and J.A. also thank GyuChul Myeong for useful discussions.
J.A. \& C.L. acknowledge funding from the European Research Council (ERC) under the European Union’s Horizon 2020 research and innovation programme (grant agreement No. 852839).
V.P.D. and L.B.S. were supported by STFC Consolidated grant \#ST/R000786/1. L.B.S acknowledges NASA-ATP award 80NSSC20K0509 and Science Foundation AAG grant AST-2009122. The simulations in this paper were run at the High Performance Computing Facility of the University of Central Lancashire.

%





\appendix\label{sec:appen}
\restartappendixnumbering

\section{APOGEE DR17 data selection}\label{appen:apogee}
We use the crossmatch of APOGEE DR17 \citep{apogeeDR17} with \textit{Gaia} EDR3 \citep{gaia1, gaiaEDR3} to select accreted star candidates based on their chemical abundances, derived from the APOGEE Stellar Parameters and Abundances Pipeline. We select only stars without flags making use of the {\tt ASPCAPFLAG} \citep{aspcap}\footnote{See Examples in \url{https://www.sdss.org/dr17/irspec/catalogs/}}. We select stars with [Al/Fe] $<-0.1$ and [Mg/Mn] $>0.2$, as the [Al/Fe]-[Mg/Mn] diagram has been shown to efficiently segregate ex-situ from in-situ stars in the MW \citep[as explored e.g. in][]{hawkins+2015, das+2020, feuillet+2021, perottoni+2021}. We also require the uncertainty in Al, Mg, Mn and Fe to be $<0.2$ dex. We opt to use a red giant sample, i.e.  stars with surface gravity $\log g < 3$, and remove stars with invalid estimations of temperatures and $\log g$. Finally, we select stars with $\varpi/\varpi_{err} > 5$,  $\varpi$ and $\varpi_{err}$ are the parallax and its error respectively, in order to have reliable estimates of their orbital parameters.\par
We calculate the position and velocities of the stars assuming the Solar Galactocentric distance $R_{\odot}=8.27$ kpc and local circular velocity $v_c = 238$ km/s \citep{schon2012akin}. We assume the Solar motion with respect to the Local Standard of Rest to be $(U_{\odot},V_{\odot}, W_{\odot}) = (11.1, 12.24, 7.25)$ km/s \citep{schon2010}. Finally, the orbits and actions were calculated with {\tt AGAMA} (details in Section \ref{sec:results}) with the gravitational potential described in \citet{mcmillan2017}.

\section{Energy-$L_z$ of additional models}

\begin{figure}
    \includegraphics[width=\textwidth]{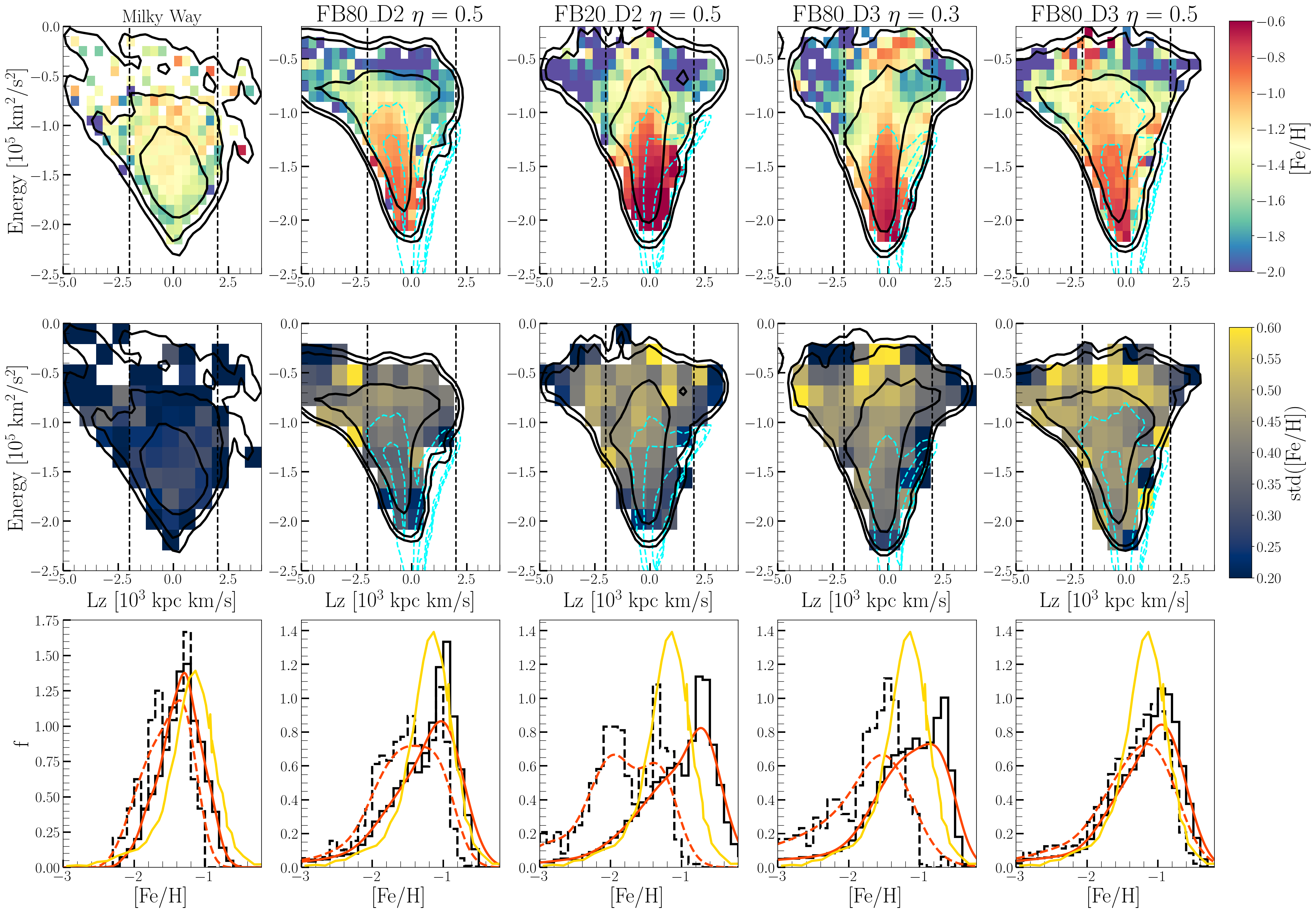}
    \caption{Same as Figure \ref{fig:enlz} but for models \fbtdt\, and \fbedt\, with $\eta=0.5$ and FB80\_D3 with $\eta=0.3$ and $0.5$.  }
    \label{fig:appen:enlz}
    \end{figure}

Figure \ref{fig:appen:enlz} presents four extra models and their [Fe/H] features in the Energy-$L_z$ plane. The second and third row show models \fbedt\, and \fbtdt\, which have the same initial conditions as the models presented in the main text, but the satellite has an initial orbit circularity of $\eta=0.5$. We also show two extra models with a dwarf \dthr, described in Section \ref{sec:dwarfIC}. The dwarfs in these models start at $r_{gc}=120$ kpc and they are run in the low feedback regime. The fourth and fifth columns of Figure \ref{fig:appen:enlz} show the cases where the initial orbit circularity of the satellite is $\eta=0.3$ and $\eta=0.5$, respectively. This shows that the $E-L_z$ space is qualitatively similar for different initial orbital circularity of the satellite. 
%

\bibliography{ref}{}
\bibliographystyle{aasjournal}



\end{document}